\documentclass[prb,aps,twocolumn,showpacs,floats]{revtex4}
\usepackage{graphicx,amsmath,epsf}

\begin{document}
\title{Towards a semiclassical justification of the `effective random 
matrix theory' for transport through ballistic chaotic quantum dots}

\author{Piet W.\ Brouwer and Saar Rahav}

\affiliation{Laboratory of Atomic and Solid State Physics, Cornell
  University, Ithaca 14853, USA}

\date{\today}

\begin{abstract}
The scattering matrix $S$ of a ballistic chaotic cavity is the direct sum
of a `classical' and a `quantum' part, which describe the scattering
of channels with typical dwell time smaller and larger than the
Ehrenfest time, respectively.  According to the `effective random
matrix theory' of Silvestrov, Goorden, and Beenakker [Phys.\ Rev.\
  Lett.\ {\bf 90}, 116801 (2003)], statistical averages involving the
quantum-mechanical scattering matrix are given by random matrix
theory. While this effective random matrix theory is known not to be
applicable for quantum interference corrections to transport, which
appear to subleading order in the number of scattering channels $N$,
it is believed to correctly describe quantum transport to leading
order in $N$. We here partially verify this belief, by comparing the 
predictions
of the effective random matrix theory for the ensemble averages of
polynomial functions of $S$ and $S^{\dagger}$ of degree $2$, $4$, and $6$
to a semiclassical calculation.
\end{abstract}


\maketitle

\section{Introduction}

For electrons moving in a ballistic conductor with chaotic classical
dynamics there is a minimal time after which the wave nature of 
electrons becomes apparent: the Ehrenfest time.\cite{kn:aleiner1996}
The Ehrenfest time $\tau_{\rm E}$ is
the time it takes for classical trajectories initially a `quantum
distance' apart 
to separate and reach a `classical 
distance'.\cite{kn:larkin1968,kn:zaslavsky1981} Here, the quantum
distance is the Fermi wavelength $\hbar/p_F$, where $p_F$ is the
Fermi momentum, whereas the classical distance can be taken to be the 
system size $L$ or the
width $W$ of contacts connecting the sample to source and drain
reservoirs. For chaotic classical dynamics with Lyapunov exponent 
$\lambda$, $\tau_{\rm E}$ is then determined by the 
condition $W = (\hbar/p_F) \exp(\lambda \tau_{\rm E})$, so that
\begin{equation}
  \tau_{\rm E} = \frac{1}{\lambda} \ln \frac{p_F W}{\hbar}.
  \label{eq:EhrDef}
\end{equation}

If the Ehrenfest time $\tau_{\rm E}$ is 
small in comparison to the mean dwell time
$\tau_{\rm D}$ in the conductor and other time scales relevant for
quantum transport (such as the dephasing time or the period of an
applied AC bias) --- which is true for most experimentally realized
ballistic quantum dots\cite{kn:kouwenhoven1997} ---, 
the time threshold it poses for quantum processes is irrelevant, 
which explains why quantum signatures do not distinguish
ballistic conductors from their disordered counterparts. Indeed,
quantum transport in ballistic quantum dots with $\tau_{\rm E} \ll
\tau_{\rm D}$ and quantum transport in disordered quantum dots are
both described by random matrix theory.\cite{kn:beenakker1997}

Recently, there has been considerable interest in the theoretical
question what happens if the Ehrenfest time exceeds the mean dwell
time $\tau_{\rm D}$. In this regime significant
differences between the manifestations of quantum mechanics in
ballistic chaotic and disordered conductors can occur.
It is important to distinguish the effect
of the Ehrenfest time on quantum phenomena that involve the splitting
of trajectories only and quantum interference
phenomena, which involve the divergence of classical trajectories
initially a quantum distance apart {\em and} their joining
again. Examples of the former are shot noise
\cite{kn:agam2000,kn:silvestrov2003,kn:tworzydlo2003,kn:whitney2005} 
and the excitation gap induced by the proximity to a 
superconductor.\cite{kn:adagideli2002,kn:vavilov2003,kn:silvestrov2003b,kn:beenakker2004}
Examples of the latter are weak 
localization,\cite{kn:aleiner1996,kn:adagideli2003,kn:rahav2005,kn:jacquod2006} 
universal conductance fluctuations,
\cite{kn:tworzydlo2004,kn:jacquod2004,kn:brouwer2006} and quantum
corrections to the level density.\cite{kn:aleiner1997,kn:tian2004b} 
Here we focus on the first type of phenomena. The effects of the
Ehrenfest time on quantum interference effects have proven quite
subtle, and will not be discussed here (see the references cited above
for details).

Silvestrov, Goorden, and Beenakker proposed a very attractive theoretical
picture to describe Ehrenfest-time related phenomena in a ballistic
chaotic quantum dot.\cite{kn:silvestrov2003b} They noted that the dot's
classical phase space can be divided into a part containing
classical trajectories with dwell time shorter than $\tau_{\rm E}$ and
a part with classical trajectories with dwell time
longer than $\tau_{\rm E}$. Correspondingly, the dot's $N \times N$
scattering
matrix $S(\varepsilon)$ is written as the direct sum of a scattering
matrix $S_{\rm cl}(\varepsilon)$ for
$N_{\rm c}$ `classical channels' and a scattering matrix $S_{\rm
  q}(\varepsilon)$ for $N_{\rm q}$ `quantum channels',
with $N = N_{\rm c} + N_{\rm q}$,
\begin{equation}
  S(\varepsilon) = \left( \begin{array}{cc} S_{\rm cl}(\varepsilon) & 0 \\ 0 & S_{\rm q}(\varepsilon)
  \end{array} \right).
  \label{eq:Sdecomp}
\end{equation}
The $N_{\rm c}$-dimensional
`classical scattering matrix' $S_{\rm cl}(\varepsilon)$ represents fully
deterministic scattering from the quantum dot, where all probability
intensity is concentrated around one classical trajectory. 
The scattering phase
shifts in $S_{\rm cl}(\varepsilon)$ follow from the dwell times of the
corresponding classical trajectories. On the other hand,
$S_{\rm q}(\varepsilon)$ represents quantum scattering, where the 
probability intensity is divided over a large number of scattering 
modes in all contacts. Writing
\begin{equation}
  S_{\rm q}(\varepsilon) = e^{i \varepsilon \tau_{\rm E}}
  \tilde S_{\rm q}(\varepsilon),
\end{equation}
in order to factor out a trivial energy dependence arising from the fact
that all trajectories contribution to $S_{\rm q}$ have a minimal dwell
time $\tau_{\rm E}$,\cite{kn:tworzydlo2004c,kn:beenakker2004}
Silvestrov, Goorden, and Beenakker proposed that 
the statistical distribution of $\tilde S_{\rm q}(\varepsilon)$
is that of random matrix theory,\cite{kn:silvestrov2003b} provided
$N_{\rm q}$ is large.

Since the original proposal of Ref.\ \onlinecite{kn:silvestrov2003b},
it has been understood that this `effective random matrix theory' does
not provide a faithful description of all signatures of quantum
transport. For example, the effective random matrix theory
predicts that the weak
localization correction to the conductance of a ballistic
quantum dot is Ehrenfest-time independent, whereas both microscopic
semiclassical theory and numerical simulations find an exponential
dependence on $\tau_{\rm E}/\tau_{\rm
  D}$.\cite{kn:aleiner1996,kn:adagideli2003,kn:rahav2005,kn:brouwer2006,kn:jacquod2006} Weak localization is a quantum interference effect, which arises as a correction to {\em subleading} order in the total channel number $N$. The effective random matrix theory has been successful in predicting and explaining Ehrenfest-time dependences that appear to leading order in $N$, such as shot noise,\cite{kn:tworzydlo2003,kn:silvestrov2003,kn:agam2000,kn:whitney2005} the proximity-induced gap in a quantum dot
coupled to a superconductor,\cite{kn:adagideli2002,kn:vavilov2003,kn:silvestrov2003b,kn:beenakker2004,kn:goorden2005} the density of transmission
eigenvalues,\cite{kn:schomerus2005} 
and the probability distribution of lifetimes of
quasibound states in chaotic quantum dots.\cite{kn:schomerus2004}

Whereas the effective random matrix theory 
has been compared to the results of accurate numerical simulations in
all four examples mentioned above, the
microscopic verification of the effective random matrix theory is
limited to a theoretical construction of the decomposition
(\ref{eq:Sdecomp})\cite{kn:jacquod2004} and to the comparison of the
effective random matrix theory and microscopic calculations
of the $\tau_{\rm E}$ dependence of the shot noise power of a chaotic 
quantum dot\cite{kn:agam2000,kn:whitney2005} and the density of states 
of a quantum dot coupled to a superconductor (an `Andreev quantum
dot') for energies much larger 
than the proximity-induced energy gap.\cite{kn:vavilov2003} (For 
energies comparable to the energy gap, the theory of Ref.\
\onlinecite{kn:vavilov2003} makes use of an ansatz that is 
similar in spirit
to the ansatz of effective random matrix theory.) The
comparison to a microscopic
calculation of shot noise amounts to a test of the effective random 
matrix theory for traces of
polynomial functions of the scattering matrix $S$ and its hermitian
conjugate $S^{\dagger}$ of degree $4$. The comparison to the density
of states of an Andreev quantum dot at high
energies addresses of the statistics of polynomials of $S_{\rm
  cl}(\varepsilon) S_{\rm cl}^{*}(-\varepsilon)$ of arbitrary degree,
but not of $S_{\rm q}$.
The aim of this article is to
provide the next step towards a microscopic verification of the
effective random matrix theory by comparing a semiclassical theory of
the ensemble average of a trace of a degree-six polynomial function of
$S$ and $S^{\dagger}$ with the predictions of the effective random
matrix theory. Since the hypothesis of the effective random matrix was
formulated after the microscopic calculations of the Ehrenfest-time
dependence of shot noise and the density of states in an Andreev
quantum dot, we believe that this calculation is the first nontrivial
test of the effective random matrix theory.


In Sec.\ \ref{sec:2} below we review the predictions of the effective
random matrix theory for the averages of traces of polynomials of $S$
and $S^{\dagger}$ of degree two, four, and six. We follow with a
semiclassical calculation of these averages in Sec.\
\ref{sec:3} and conclude in Sec.\ \ref{sec:4}. Details of the
calculations can be found in the two appendices.

\bigskip

\section{Predictions of the effective random matrix theory}
\label{sec:2}

In our calculation, we consider a ballistic quantum dot coupled to
electron reservoirs through ballistic point contacts. Hence, the
scattering matrix $S$ acquires a block structure $S = S_{ij}$, where
the indices $i$ and $j$ label the point contacts. The dimension of
the block $S_{ij}$ is $N_i \times N_j$, where $N_i$ is the number of
channels in the $i$th point contact. Each block $S_{ij}$ can be
decomposed into a `classical' and a `quantum' scattering matrix as in
Eq.\ (\ref{eq:Sdecomp}). 

We are interested in averages of the form
\begin{eqnarray}
  \label{eq:Adef}
  Q_2 &=& \frac{1}{N}
  \langle \mbox{tr}\, S_{ij}^{\vphantom{\dagger}}(\varepsilon_1)
  S_{ij}^{\dagger}(\varepsilon_2) \rangle, \nonumber \\
  Q_4 &=&  \frac{1}{N}
 \langle \mbox{tr}\, S_{ij}^{\vphantom{\dagger}}(\varepsilon_1)
  S_{kj}^{\dagger}(\varepsilon_2) S_{kl}^{\vphantom{\dagger}}(\varepsilon_3) S_{il}^{\dagger}(\varepsilon_4)
  \rangle, \nonumber \\
  Q_6 &=&  \frac{1}{N}
 \langle \mbox{tr}\, S_{ij}^{\vphantom{\dagger}}(\varepsilon_1)
  S_{kj}^{\dagger}(\varepsilon_2) 
  S_{kl}^{\vphantom{\dagger}}(\varepsilon_3)
  \ldots
  S_{in}^{\dagger}(\varepsilon_6) \rangle.
\end{eqnarray}
The polynomials $Q_2$ and $Q_4$ describe, {\em e.g.}, the conductance and
shot noise power of a quantum dot with normal-metal 
contacts,\cite{kn:blanter2000b} whereas polynomials of higher degree
are necessary for a theory of transport and equilibrium
properties of quantum dots with superconducting
contacts.\cite{kn:beenakker2004}
According to the effective random matrix theory, these averages have
the structure
\begin{eqnarray}
  \label{eq:effrmt}
  Q_n &=& \left[1 - e^{-F_n(1,\ldots,n)\tau_{\rm E}/\tau_{\rm D}}
  \right]
  Q_n^{\rm cl} \nonumber \\ && \mbox{} + 
  e^{-F_n(1,\ldots,n)\tau_{\rm E}/\tau_{\rm D}} Q_n^{\rm RMT},
\end{eqnarray}
where the function $F_n$ is defined as
\begin{equation}
  F_n(1,\ldots,n) = 1 + \frac{i \tau_{\rm D}}{\hbar}
  \sum_{j=1}^{n} (-1)^j \varepsilon_j,
  \label{eq:F}
\end{equation}
and $Q^{\rm cl}$ and $Q^{\rm RMT}$ are obtained from Eq.\
(\ref{eq:Adef}) by replacing $S$ by $S_{\rm cl}$ and $\tilde S_{\rm
q}$ and $N$ by $N_{\rm c}$ and $N_{\rm q}$, respectively.

The classical averages $Q_n^{\rm cl}$ read\cite{kn:silvestrov2003b}
\begin{eqnarray}
  \label{eq:Acl}
  Q_n^{\rm cl} &=&
  \frac{N_i N_j}{N^2 F_2(1,2)}.
  \nonumber \\
  Q_4^{\rm cl} &=&
  \frac{N_i N_j \delta_{ik} \delta_{jl}}{N^2 F_4(1,2,3,4)}, \nonumber \\
  Q_6^{\rm cl} &=&
  \frac{N_i N_j \delta_{ik} \delta_{jl} \delta_{im} \delta_{jn}}{N^2 F_6(1,2,3,4,5,6)}.
\end{eqnarray}
The random matrix average of the trace of a degree two
polynomial of $S$ and $S^{\dagger}$ is equal to the classical
average,\cite{kn:bluemel1988}
\begin{equation}
  \label{eq:Armt1}
  Q_2^{\rm RMT} = Q_2^{\rm cl}.
\end{equation}
However, the higher-order random matrix
averages are different. For $Q_4$ one 
has\cite{kn:polianski2003}
\begin{widetext}
\begin{eqnarray}
  \label{eq:Armt2}
  Q_4^{\rm RMT} &=&
  \frac{N_i N_j N_l \delta_{ik}}{N^3 F_2(1,2) F_2(3,4)} +
  \frac{N_i N_j N_k \delta_{jl}}{N^3 F_2(3,2) F_2(1,4)} 
  - 
  \frac{N_i N_j N_k N_l F_4(1,2,3,4)}{N^4 F_2(1,2) F_2(3,2) F_2(3,4)
  F_2(1,4)},
\end{eqnarray}
whereas the random matrix average $Q_6$ 
reads\cite{kn:brouwer2005d}
\begin{eqnarray}
  \label{eq:Armt3}
  Q_6^{\rm RMT} &=&
  \frac{N_i N_j N_l N_n \delta_{ik} \delta_{im}}{N^4 F_2(1,2) F_2(3,4)
  F_2(5,6)} +
  \frac{N_i N_j N_k N_m \delta_{jl} \delta_{jn}}{N^4 F_2(3,2) F_2(5,4)
  F_2(1,6)}
  -
  \frac{N_i N_j N_k N_l N_m N_n F_6(1,2,3,4,5,6)}{
  F_2(1,2) F_2(3,2) F_2(3,4) F_2(5,4) F_2(5,6) F_2(1,6)}
   \nonumber \\ && \mbox{} + 
  \left[
  \frac{N_i N_j N_l N_m \delta_{ik} \delta_{ln}}{N^4 F_2(1,2) F_2(3,6)
  F_2(5,4)} 
  +  \frac{N_i N_j N_k N_l N_m N_n F_4(1,2,3,4) F_4(1,4,5,6)}{
  F_2(1,2) F_2(3,2) F_2(3,4) F_2(5,4) F_2(5,6) F_2(1,6) F_2(1,4)}
  \right. \nonumber \\ && \left. \mbox{}  
  -
  \frac{N_i N_j N_l N_m N_n F_4(3,4,5,6) \delta_{ik}}{N^5 F_2(1,2)
  F_2(3,4) F_2(5,4) F_2(5,6) F_2(3,6)} -
  \frac{N_i N_j N_k N_m N_n F_4(1,4,5,6) \delta_{jl}}{N^5 F_2(3,2)
  F_2(1,4) F_2(5,4) F_2(5,6) F_2(1,6)} 
  + \ldots
  \right],
\end{eqnarray}  
\end{widetext}
where the dots $\ldots$ refer to the simultaneous 
cyclic permutations $(1,2,i,j) \to (3,4,k,l) \to (5,6,m,n)$.

Together, Eqs.\ (\ref{eq:effrmt}), (\ref{eq:Acl}),
(\ref{eq:Armt1}), (\ref{eq:Armt2}), and (\ref{eq:Armt3}) specify the
prediction of the effective random matrix theory for the averages of
the traces of polynomials of the scattering matrix and its hermitian
conjugate for polynomials up to degree 6.


\section{Semiclassical calculation} 
\label{sec:3}

We now perform a semiclassical calculation of the
averages $Q_2$, $Q_4$, and $Q_6$ defined in Eq.\
(\ref{eq:Adef}) above and show that they agree with the predictions of
the effective random matrix theory as described in the previous
section.
Although semiclassical calculations of $Q_2$ and $Q_4$ exist, 
see Refs.\
\onlinecite{kn:bluemel1988,kn:jalabert1990,kn:baranger1993,kn:baranger1993b}
and
\onlinecite{kn:agam2000,kn:whitney2005,kn:braun2005}, respectively, we  
briefly review their derivation in order to establish the
context for the calculation of $Q_6$. (We are not aware of a
semiclassical calculation of the energy dependence of $Q_4$, however.)

In our calculation, we take the limit 
$\hbar \to 0$ while keeping the ratios $\tau_{\rm E}/\tau_{\rm D}$ and
$N_i/N$, 
and the products $\tau_{\rm D} \varepsilon_i/\hbar$ fixed. The latter
condition implies that the functions $F_n$ defined in Eq.\
(\ref{eq:F}) remain constant in the limiting procedure.
Note that the channel numbers $N_i$ diverge in this limit. The 
divergence of the channel numbers does not affect
 $Q_2$, $Q_4$, or $Q_6$, however, because these depend on the ratios
$N_i/N$ only. Since $\tau_{\rm E} \propto \ln (1/\hbar)$, see Eq.\
(\ref{eq:EhrDef}), the condition that
the ratio $\tau_{\rm E}/\tau_{\rm D}$ is kept constant in the limiting
procedure implies that the dwell time $\tau_{\rm D}$ diverges as well.
The divergence of  $\tau_{\rm D}$ removes
any dependence on the non-universal short-time dynamics in the quantum
dot.

\begin{figure}
\epsfxsize=0.7\hsize
\hspace{0.05\hsize}
\epsffile{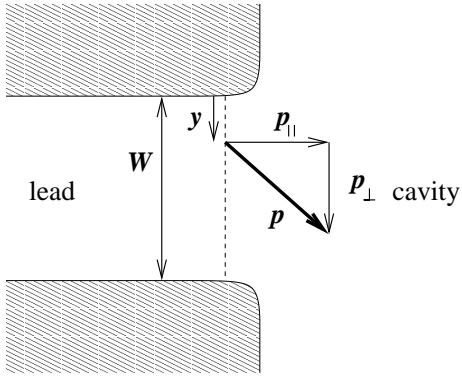}
\caption{\label{fig:contact}
Schematic drawing of one of the contacts to the quantum dot. The coordinate
$y$ as well as the perpendicular component of the momentum $p_{\perp}$ 
refers to the direction perpendicular to the lead axis. In the sum
(\ref{eq:Ssemi0}) over classical trajectories there is no restriction
on $y$, but only discrete values of $p_{\perp}$ are considered.}
\end{figure}

Starting point of our calculation 
is an expression of the dot's scattering matrix $S$ as
a sum over classical trajectories 
$\alpha$,\cite{kn:jalabert1990,kn:baranger1993,kn:baranger1993b} 
\begin{equation}
  (S_{ij})_{mn} = 
  \left( \frac{\pi \hbar}{2 W_i W_j} \right)^{1/2}
  \sum_{\alpha} A_{\alpha} e^{i {\cal S}_{\alpha}/\hbar},
  \label{eq:Ssemi0}
\end{equation}
where $i$ and $j$ label the exit and entrance leads, respectively, and
$m$ and $n$ label the propagating modes in these leads. 
The widths of the entrance and exit contacts are $W_j$ and $W_i$, 
respectively. The trajectory
$\alpha$ connects the entrance contact $j$ to the exit contact
$i$.
The components $p_{\perp}$ and $p_{\perp}'$ of its momentum 
perpendicular to the lead axis upon
entrance and exit, respectively, are compatible
with that of the modes $n$ and $m$ in the corresponding leads,
\begin{eqnarray}
  \label{eq:pW}
  p_{\perp} &=&
  \pm \pi \hbar n/W_{j},\ \ n=1,\ldots,N_{j},
  \nonumber \\
  p_{\perp}'
  &=& \pm \pi \hbar m/W_{i},\ \ m=1,\ldots,N_{i}.
\end{eqnarray}
(Here and in the
remainder of this article, primed variables refer to the exit
contact.)
Further, ${\cal S}_{\alpha}$ is the classical action of trajectory
$\alpha$ and ${A}_{\alpha}$ is its stability amplitude. The latter
is defined as
\begin{equation}
  A_{\alpha} =
  \left| \frac{\partial p_{\perp}'}{dy} \right|^{-1/2}
  \label{eq:Aampl},
\end{equation}
where $y$ is the coordinate perpendicular to the axis of the
entrance contact, see Fig.\ \ref{fig:contact}, and 
the partial derivative is taken at constant $p_{\perp}$. 
The classical action ${\cal S}_{\alpha}$ 
is a function of $p_{\perp}$ and $p_{\perp}'$. 
The spatial coordinates
$y$ and $y'$ of the trajectory $\alpha$ upon entrance and exit
can be expressed as derivatives of ${\cal S}_{\alpha}$,
\begin{equation}
  y = \frac{\partial {\cal S}_{\alpha}}{\partial p_{\perp}}, \ \
  y' = - \frac{\partial {\cal S}_{\alpha}}{\partial p_{\perp}'}.
\end{equation}
For simplicity of
notation, the Maslov index and other phase shifts are included in 
${\cal S}_{\alpha}$. 

We are interested in the scattering matrix $S_{ij}$ at different
values of the energy $\varepsilon$. The energy $\varepsilon$ enters in
Eq.\ (\ref{eq:Ssemi0}) through the energy dependence of the action
${\cal S}_{\alpha}$,
\begin{equation}
  {\cal S}_{\alpha}(\varepsilon) = {\cal S}_{\alpha}(\varepsilon_0)
  + (\varepsilon - \varepsilon_0) t_{\alpha},
\end{equation}
where $\varepsilon_0$ is a reference energy and $t_{\alpha}$ is the
duration of the trajectory $\alpha$. We neglect the energy dependence
of the stability amplitudes $A_{\alpha}$.

Using Eq.\ (\ref{eq:Ssemi0}), the 
traces $Q_n$, $n=2,4,6$, are expressed as double, quadruple,
and sixfold summations over classical trajectories,
\begin{eqnarray}
  \label{eq:Q2sum}
  Q_2 &=&
  \frac{\pi \hbar}{2 W_i W_j} \sum_{\alpha_1,\alpha_2}
  A_1 A_2 e^{i({\cal S}_1 - {\cal S}_2)/\hbar},
\end{eqnarray}
with similar expressions for $Q_4$ and $Q_6$.
Here and in the remainder of this article we use the indices $1$ and $2$ to denote
$\alpha_1$ and $\alpha_2$ if possible without 
confusion.
The transverse momentum components $p_{\perp}$ and
$p_{\perp}'$ at the entrance and exit contacts of the $n$ trajectories
involved in the trajectory sum for $Q_n$ satisfy
\begin{eqnarray}
  p_{\perp,1} &=& p_{\perp,2}, \ \ \ldots,\ \
  p_{\perp,n-1} = p_{\perp,n}, \nonumber \\
  p_{\perp,2}' &=& p_{\perp,3}',\ \ \ldots,\ \ p_{\perp,n}' =
  p_{\perp,1}',
  \label{eq:pcond}
\end{eqnarray} 
where $n=2,4,6$.\cite{foot}
In the summation (\ref{eq:Q2sum}), the magnitudes of all transverse 
momentum components are taken 
equal to the quantized values (\ref{eq:pW}). However, for large
channel numbers we may replace the summation over the quantized
momenta by integrations over $p_{\perp}$ and $p_{\perp'}$, so that
\begin{eqnarray}
  Q_2 &=& \frac{1}{2 \pi \hbar}
  \int dp_{\perp,1} dp_{\perp,1}'
  \sum_{\alpha_1,\alpha_2}
  A_1 A_2 e^{i({\cal S}_1 - {\cal S}_2)/\hbar},~~~
  \label{eq:Qn}
\end{eqnarray}
again with similar expressions for $Q_4$ and $Q_6$. In Eq.\
(\ref{eq:Qn}) and its equivalents for $Q_4$ and $Q_6$, one integrates
over the perpendicular momentum components $p_{\perp}$ and
$p_{\perp}'$
of the trajectories
$\alpha_1,\ldots,\alpha_{n-1}$ with odd indices. The 
initial and final transverse momentum components of the
classical trajectories $\alpha_2, \ldots, \alpha_{n}$ with even
indices are determined by the conditions (\ref{eq:pcond}).

For the calculations below, we assume that the classical dynamics
inside the quantum dot is uniformly hyperbolic. This assumption
simplifies the calculations, although it is believed not to affect the
final results in the universal limit $\tau_{\rm D} \gg
L/v_F$.\cite{kn:mueller2004,kn:mueller2005,kn:heusler2006}
With uniformly hyperbolic classical dynamics, the optimal
phase space coordinates for a Poincar\'e surface of section taken in
the interior of the quantum dot are coordinates $s$ and $u$ taken 
along the stable and unstable directions in phase space. In view of 
this, we replace the trajectory sum (\ref{eq:Qn}), which is taken over
trajectories with specified transverse momentum components $p_{\perp}$
and $p_{\perp}'$ at entrance and exit contacts, by a trajectory sum
over trajectories with specified stable and unstable phase space
coordinates $s$ and $u'$ at entrance and exit contacts,
respectively. Referring to App.\ \ref{app:A} for details, we find
that the traces $Q_2$, $Q_4$, and $Q_6$ can also be expressed as
\begin{eqnarray}
  Q_2 &=& \frac{1}{2 \pi \hbar}
  \int ds_1 du_1'
  \sum_{\alpha_1,\alpha_2}
  A_1 A_2 e^{i({\cal S}_1 - {\cal S}_2)/\hbar},
  \label{eq:Qn2}
  \\
  Q_4 &=& \frac{1}{(2 \pi \hbar)^2}
  \int ds_1 du_1' ds_3 du_3' 
  \nonumber \\ && \mbox{} \times 
  \sum_{\alpha_1,\ldots,\alpha_4} 
  A_1 \ldots A_4
  e^{i({\cal S}_1 - \ldots - {\cal S}_4)/\hbar},
  \label{eq:Qn4} \\
  Q_6 &=& \frac{1}{(2 \pi \hbar)^3}
  \int ds_1 du_1' ds_3 du_3' ds_5 du_5'
  \nonumber \\ && \mbox{} \times 
  \sum_{\alpha_1,\ldots,\alpha_6} 
  A_1 \ldots A_6
    e^{i({\cal S}_1 - \ldots - {\cal S}_6)/\hbar},
  \label{eq:Qn6}
\end{eqnarray}
where the classical trajectories $\alpha_{\mu}$, $\mu=1,\ldots,n$
satisfy the conditions
\begin{eqnarray}
  s_1 &=& s_2, \ \ \ldots,\ \ s_{n-1} = s_n, \nonumber \\
  u_2' &=& u_3',\ \ \ldots,\ \ u_{n}' = u_1',
  \label{eq:pcondsu}
\end{eqnarray}
with $n=2,4,6$. Further, in Eqs.\ (\ref{eq:Qn2})--(\ref{eq:Qn6}),
the stability amplitudes are defined as 
\begin{equation}
  A_{\alpha} = \left| \frac{\partial u'}{\partial u}
  \right|^{-1/2},
  \label{eq:Asu}
\end{equation}
and the classical actions ${\cal S}_{\alpha}(s,u')$ are Legendre transforms of
the original classical actions ${\cal S}_{\alpha}(p_{\perp},p_{\perp}')$, so
that
\begin{equation}
  \frac{\partial {\cal S}_{\alpha}}{\partial s} =
  u_{\alpha}, \ \
  \frac{\partial {\cal S}_{\alpha}}{\partial u'} = -s_{\alpha}'.
\end{equation}

For the interpretation of Eqs.\ (\ref{eq:Qn2})--(\ref{eq:Qn6}), one
should keep in mind that the phase space coordinates $s$ and $u$ are
defined only locally. That means that the coordinate
transformation $(x,p_{\perp}) \to (s,u)$ can only be made for pairs of
trajectories that enter or exit the quantum dot at nearby phase space
points. This poses no problems for our calculation, because only 
classical trajectories that exit the quantum dot at nearby 
positions and with close momenta have an action difference $\Delta 
{\cal S}$ that varies sufficiently slowly as a function of the phase 
space coordinates to give a finite contribution to $Q_n$, 
$n=2,4,6$. Indeed, the explicit calculations of $Q_2$, $Q_4$, and
$Q_6$ in the following subsections show
that the entire contribution to $Q_n$ comes from trajectories for
which the differences of the stable or unstable phase space
coordinates are small.

\subsection{Calculation of $Q_2$}
\label{sec:3a}

The leading contribution to $Q_2$ arises from equal trajectories
$\alpha_1 = \alpha_2$. In that case, the action difference 
\begin{eqnarray}
  \Delta {\cal S} &=&
  {\cal S}_{1} - {\cal S}_{2} \nonumber \\
  &=& (\varepsilon_1 - \varepsilon_2) t,
\end{eqnarray}
where $t$ is the common duration of the trajectories $\alpha_1$ and 
$\alpha_2$. The stability amplitudes are equal, $A_1 = A_2 = A$.
The factor $A_1 A_2 = A^2$ in Eq.\ (\ref{eq:Qn2}) 
provides the Jacobian necessary to replace the integration over the
unstable phase space coordinate $u'$ at the exit contact by an
integration over the unstable phase space coordinate $u$ at the
entrance contact.\cite{kn:jalabert1990,kn:baranger1993,kn:baranger1993b}
This way, $Q_2$ is expressed in terms of an integration over the phase
space coordinates $s$ and $u$ at a Poincar\'e surface of
section taken in the entrance contact.
Equivalently, the double integral over $s$ and $u'$
can be represented by a phase space integral over any Poincar\'e
surface of section taken anywhere along the classical trajectory
$\alpha_1=\alpha_2$.\cite{kn:heusler2006} Following the latter
strategy, we write $Q_2$ as
\begin{equation}
  Q_2 =
  \int dq \int_0^{\infty} dt_1 dt_2
  \frac{P_i(t_1) P_j(t_2) e^{i
  (\varepsilon_1-\varepsilon_2)(t_1+t_2)/\hbar}}{2 \pi \hbar
  (t_1+t_2) N},
  \label{eq:Q21}
\end{equation}
where $q$ refers to the phase space coordinate at which the reference
surface of section is taken, $P_i(t_1)$ selects only those $q$
for which the classical propagation of $q$ ends up in contact $i$
after time $t_1$, and $P_j(t_2)$ selects $q$
for which the classical propagation of the time-reversed of $q$ ends
at contact $j$ after a time $t_2$. We divided by $t = t_1
+ t_2$
to cancel a spurious contribution from the freedom to choose the 
reference surface
of section at an arbitrary point along the trajectory
$\alpha_1=\alpha_2$. Replacing $P_i$ and $P_j$ by classical probabilities to
reach the contacts $i$ and $j$ for an arbitrary phase space point $q$,
\begin{equation}
  P_i(t) = \frac{N_i}{N \tau_{\rm D}} e^{-t/\tau_{\rm D}},\ \
  P_j(t) = \frac{N_j}{N \tau_{\rm D}} e^{-t/\tau_{\rm D}},
  \label{eq:Pij}
\end{equation}
the integration over $q$ contributes the total phase space volume $2
\pi \hbar N \tau_{\rm D}$ of the quantum dot and we find the
well-known result
\begin{equation}
  Q_2 = \frac{N_i N_j}{N^2 F_2(1,2)}
\end{equation}
upon integration over $t_1$ and $t_2$.

\subsection{Calculation of $Q_4$}
\label{sec:3b}

\begin{figure}
\epsfxsize=0.95\hsize
\epsffile{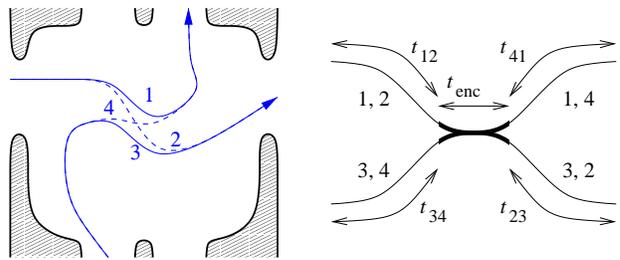}
\caption{\label{fig:1} Left: Four classical
  trajectories that give the random matrix contribution to the average
  $Q_4$. 
  These trajectories have a small-angle encounter which
  fully resides inside the quantum dot. Right: Schematic drawing of
  the encounter together with the definitions of the various times
  used in the text. The encounter region is shown thick.}
\end{figure}
\begin{figure}
\epsfxsize=0.95\hsize
\epsffile{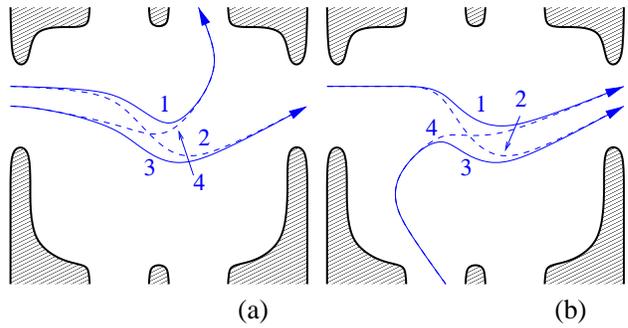}
\caption{\label{fig:1b} Schematic drawing of four classical
  trajectories that give the random matrix contribution with one
  Kronecker delta to the average
  $Q_4$. These trajectories have a small-angle encounter touches
  one of the lead openings. The Kronecker delta appears for the
  entrance lead index in (a) and for the exit lead index in (b).}
\end{figure}
\begin{figure}
\epsfxsize=0.45\hsize
\hspace{0.05\hsize}
\epsffile{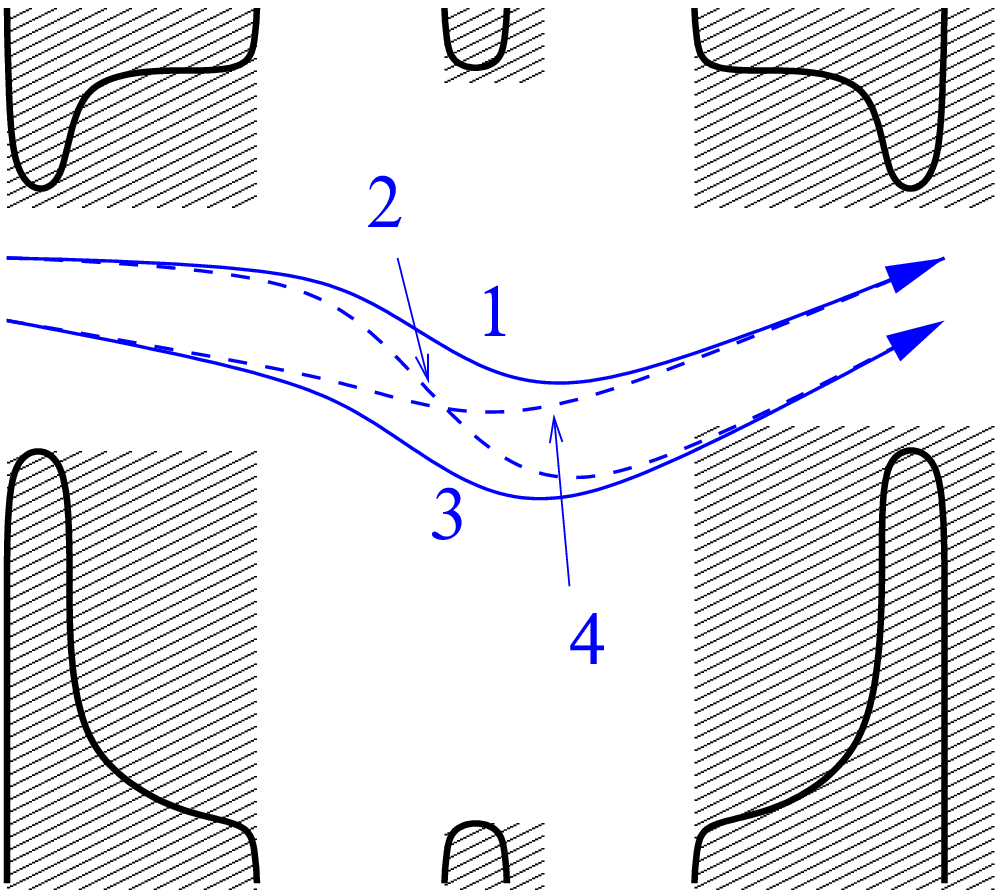}
\caption{\label{fig:0} Schematic drawing of four classical 
  trajectories with a
  small-angle encounter that touches both lead openings. 
  Such configurations of classical trajectories give
  the classical contribution to $Q_4$.}
\end{figure}

The trace $Q_4$ is expressed as a quadruple sum over classical
trajectories, see Eq.\ (\ref{eq:Qn4}). The typical configuration of 
four trajectories that contributes to $Q_4$ was pointed out in
Refs.\ \onlinecite{kn:whitney2005,kn:braun2005}: The trajectories
$\alpha_1$, $\alpha_2$, $\alpha_3$, and $\alpha_4$, which are
paired close to entrance and exit, have a small-angle 
encounter at which the pairing of the trajectories is interchanged, see 
Fig.\ \ref{fig:1}. Before arriving at the encounter, $\alpha_1$ and
$\alpha_2$, and $\alpha_3$ and $\alpha_4$ are paired. After the
encounter, $\alpha_2$ and $\alpha_3$, and $\alpha_4$ and $\alpha_1$
are paired. The
encounter may reside fully inside the quantum dot, as in Fig.\
\ref{fig:1}, so that the
pairs of trajectories are uncorrelated when they exit and
enter the quantum dot, or the small-angle
encounter may touch one or two of the lead openings, as in Figs.\
\ref{fig:1b} and \ref{fig:0}, respectively, so that exit or entrance
of the four trajectories is correlated.
While the importance of such small-angle encounters of 
classical trajectories was first realized for weak 
localization,\cite{kn:aleiner1996,kn:richter2002} they play a crucial
role in all semiclassical theories of quantum transport.\cite{kn:heusler2006,kn:braun2005,kn:whitney2005,kn:jacquod2006,kn:brouwer2006,kn:adagideli2003}

Since all four trajectories are
fixed once we have specified $\alpha_1$ and $\alpha_3$, we need to sum
over $\alpha_1$ and $\alpha_3$ only. In order to 
parameterize $\alpha_1$ and $\alpha_3$ we take a Poincar\'e 
surface of section chosen at an arbitrary point during the encounter. 
We then use the coordinate $q$ of a reference point at the Poincar\'e
surface of section, as well as the differences $s_3-s_1$ and $u_3-u_1$ 
of the stable and unstable phase space coordinates to parameterize
$\alpha_1$ and $\alpha_3$.\cite{kn:spehner2003}
The trajectories $\alpha_1$ and
$\alpha_2$, and $\alpha_3$ and $\alpha_4$
originate from the same contact and with the same
stable phase stable space coordinate, hence
\begin{equation}
  s_1 = s_2, \ \ s_3 = s_4.
  \label{eq:scond}
\end{equation}
Similarly,
\begin{equation}
  u_4 = u_1, \ \ u_2 = u_3.
  \label{eq:ucond}
\end{equation}
Since the trajectory pairs have the same stable and unstable phase
space coordinates at the entrance and exit contacts, respectively, the
conditions (\ref{eq:scond}) and (\ref{eq:ucond}) hold for encounters
that do not touch the contacts as well as for encounters that touch
the contacts.

We first discuss the case of Fig.\ \ref{fig:1}, in which the 
encounter resides fully inside the quantum dot. 
We closely follow Ref.\ \onlinecite{kn:braun2005}, in
which the same configuration of trajectories was considered in the
limit $\tau_{\rm E}/\tau_{\rm D} \to 0$. 
The encounter region is defined as the
segment of the trajectories for which $|s_1-s_3| < c$ and $|u_1-u_3| <
c$, where $c \sim (p_F W)^{1/2}$ 
is a classical cut-off below which the classical
dynamics can be linearized. 
The precise choice of $W$ is irrelevant in the classical
limit, see the discussion following Eq.\
(\ref{eq:A4res}). Hence, the duration of the encounter is
\begin{equation}
  t_{\rm enc} =
  \frac{1}{\lambda} \ln \frac{c^2}{|(u_3-u_1)(s_3- s_1)|},
\end{equation}
where $\lambda$ is the Lyapunov exponent for the classical dynamics in
the quantum dot.
We parameterize the durations $t_{\alpha_1}$, $t_{\alpha_2}$, $t_{\alpha_3}$,
and $t_{\alpha_4}$ of the four trajectories involved
using $t_{\rm enc}$ and
the durations $t_{12}$, $t_{23}$, $t_{34}$, and $t_{41}$ 
of the four stretches connecting the encounter region with the lead 
openings,
\begin{eqnarray}
  t_{\alpha_1} &=& t_{\rm enc} + t_{12} + t_{41}, \nonumber \\
  t_{\alpha_2} &=& t_{\rm enc} + t_{12} + t_{23}, \nonumber \\
  t_{\alpha_3} &=& t_{\rm enc} + t_{34} + t_{23}, \nonumber \\
  t_{\alpha_4} &=& t_{\rm enc} + t_{34} + t_{41}.
\end{eqnarray}
Schematically, the definitions of $t_{\rm enc}$ and of the times
$t_{12}$, $t_{23}$, $t_{34}$, and $t_{41}$ are shown in the right
panel of Fig.\ \ref{fig:1}.
The action difference $\Delta {\cal S}$ 
is the symplectic area enclosed by the four
trajectories,\cite{kn:spehner2003,kn:turek2003} plus a contribution
from the energy differences,
\begin{eqnarray}
  \label{eq:dS4}
  \Delta {\cal S} &=& (s_3-s_1)(u_3-u_1)
  \nonumber \\ && \mbox{}
  + t_{\rm enc}(\varepsilon_1 - \varepsilon_2 + \varepsilon_3 -
  \varepsilon_4) 
  \nonumber \\ && \mbox{}
  + t_{12}(\varepsilon_1 - \varepsilon_2) 
  + t_{23}(\varepsilon_3 - \varepsilon_2)
  \nonumber \\ && \mbox{}
  + t_{34}(\varepsilon_3 - \varepsilon_4)
  + t_{41}(\varepsilon_1 - \varepsilon_4).~~
\end{eqnarray}
Note that the enclosed phase space areas are
conserved along the motion of the trajectories. In particular, this
means that the action difference $\Delta {\cal S}$ is independent of where the
Poincar\'e surface of section is chosen.


Integrating over the position of the Poincar\'e surface of section, we
then find
\begin{eqnarray}
  \label{eq:P42}
  Q_4^{(\ref{fig:1})}
  &=&
  \int dt_{12} dt_{23} dt_{34} t_{41} P_{j}(t_{12}) P_{k}(t_{23})  
  P_{l}(t_{34}) P_{i}(t_{41})
  \nonumber \\ && \mbox{} \times
  \int d(u_3-u_1) d(s_3-s_1)
  \frac{\tau_{\rm D} e^{i \Delta {\cal S}/\hbar-t_{\rm enc}/\tau_{\rm
  D}}}{2 \pi \hbar t_{\rm enc}}. \nonumber \\
\end{eqnarray}
In this equation, the factor $t_{\rm enc}$ in the denominator cancels
a spurious contribution to the
integral from the freedom to choose the Poincar\'e surface of section
anywhere along the encounter region. The classical probabilities
$P_i(t_{41})$, $P_j(t_{12})$, $P_k(t_{23})$, and
$P_l(t_{34})$ are defined as in Eq.\ (\ref{eq:Pij}).
The factor $\exp(-t_{\rm enc}/\tau_{\rm D})$ is
the probability that the trajectories do not exit the quantum dot
during the encounter stretch. (If they do, the encounter touches the
lead opening. This case is treated separately below.)

Taking $u_3-u_1$ to be positive (while adding a factor $2$
in Eq.\ (\ref{eq:P42})), we perform the variable change 
\begin{equation} 
  u_3-u_1 = c/\sigma,\ \  s_3-s_1 = c x \sigma.
  \label{eq:varchg1}
\end{equation}
With the new integration variables, the integration domain is $-1 < x <
1$ and $1 < \sigma < 1/|x|$. Further, 
$t_{{\rm enc}} = \lambda^{-1} \ln(1/|x|)$. 
Integrating over $t_{12}$,
$t_{23}$, $t_{34}$, and $t_{41}$, and $\sigma$, and performing a
partial integration to $x$, we then find
\begin{eqnarray}
  Q_4^{(\ref{fig:1})} &=&  
  - \frac{N_i N_j N_k N_l F_4(1,2,3,4)}{N^4
  F_2(1,2) F_2(3,2) F_2(3,4) F_2(1,4)}
  \nonumber \\ && \mbox{} \times
  \int_{0}^{1} dx 
  x^{F_4(1,2,3,4)/\lambda \tau_{\rm D}} 
  \frac{2 \sin(x r)}{\pi x},
\end{eqnarray}
where we omitted a term that is an oscillating function of $r =
c^2/\hbar$ in the limit $\hbar \to 0$. The remaining integral over $x$
can be evaluated in the limit $\hbar \to 0$ at fixed $\tau_{\rm
  E}/\tau_{\rm D}$, and we arrive at the final result
\begin{eqnarray}
  Q_4^{(\ref{fig:1})} &=& -\frac{N_i N_j N_k N_l F_4(1,2,3,4)}{N^4
  F_2(1,2) F_2(3,2) F_2(3,4) F_2(1,4)}
  \nonumber \\ && \mbox{} \times
  e^{-\tau_{\rm E} F_4(1,2,3,4)/\tau_{\rm D}},
  \label{eq:A4res}
\end{eqnarray}
where the Ehrenfest time $\tau_{\rm E}$ is defined as, {\em cf.} Eq.\
(\ref{eq:EhrDef}),
\begin{equation}
  \tau_{\rm E} = \frac{1}{\lambda} \ln r =
  \frac{1}{\lambda} \ln \frac{c^2}{\hbar}.
\end{equation}

It is important to notice that the precise value of the phase-space
cut-off $c$ enters {\em only} through the Ehrenfest time $\tau_{\rm
  E}$. However, since the Ehrenfest time appears in the combination
$\tau_{\rm E}/\tau_{\rm D}$ only and since $\tau_{\rm D} \to \infty$
in the classical limit taken here, any change of $c$ by a factor of 
order unity will not affect the final result. It is because of this
that we did not need to carefully specify the classical length scale
$W$ entering into the cut-off $c$.

If the encounter region touches one of the lead openings, as in Figs.\
\ref{fig:1b}a and b, a slight modification of the above calculation is
called for. For
definiteness, we consider an encounter that touches the entrance lead
opening, as shown in Fig.\ \ref{fig:1b}a. This implies $j=l$, so that
the corresponding contribution to $Q_4$ is proportional to
$\delta_{jl}$. The action difference $\Delta {\cal S}$ is given by
Eq.\ (\ref{eq:dS4}) with $t_{12} = t_{34} = 0$.
Since the trajectories $\alpha_1$ and $\alpha_3$ must be
correlated upon entry for the configuration shown in Fig.\
\ref{fig:1b}a, the encounter time $t_{\rm enc}$ is bounded by
\begin{equation}
  0 < t_{\rm enc} - \frac{1}{\lambda} \ln \frac{c}{|u_3-u_1|} <
  \frac{1}{\lambda} \ln \frac{c}{|s_3-s_1|}.
\end{equation}
Then, proceeding as in the calculation of $Q_4^{(\ref{fig:1})}$ and 
integrating over $t_{23}$ and $t_{41}$, we find
\begin{eqnarray}
  Q_4^{(\ref{fig:1b}a)} &=&
  \frac{N_i N_k \tau_{\rm D} \delta_{jl}}{N^2
  F_2(3,2) F_2(1,4)} 
  \int dt_{\rm enc}
  P_{j}(t_{\rm enc})
  \nonumber \\ && \mbox{} \times
  \int_{-c}^{c}   \frac{d(s_1-s_3) d(u_1-u_3)}{2 \pi \hbar t_{\rm enc}}
  \nonumber \\ && \mbox{} \times
  e^{i ( s_3-s_1)( u_3-u_1)/\hbar + i t_{\rm enc}
  (\varepsilon_1-\varepsilon_2+\varepsilon_3-\varepsilon_4))/\hbar}.
  ~~~~
  \label{eq:Ri2sub}
\end{eqnarray}
In order to perform the integrations in Eq.\
(\ref{eq:Ri2sub}), we take $u_3-u_1$ to be positive and perform the 
variable change 
\begin{equation} 
  u_3-u_1 = \frac{c}{\sigma},\ \  s_3-s_1 = c x \sigma.
  \label{eq:varchg2}
\end{equation}
With the new integration variables, the integration domain is $-1 < x <
1$, $1 < \sigma < e^{\lambda t_{\rm enc}}$, and $0 < t_{\rm enc} < \lambda^{-1}
\ln(1/|x|)$. Hence, with 
$r = c^2/\hbar$ and integrating over $t_{\rm enc}$ and $\sigma$, 
the integral (\ref{eq:Ri2sub}) becomes
\begin{eqnarray}
  Q_4^{(\ref{fig:1b}a)} &=&
  \frac{N_i N_j N_k \delta_{jl}}{N^3 F_2(3,2) F_2(1,4)} 
  \nonumber \\ && \mbox{} \times
  \int_{0}^{1} dx \frac{2 \sin(r x)}{\pi x}
  x^{F_4(1,2,3,4)/\lambda \tau_{\rm D}} \nonumber \\
  &=&
  \frac{N_i N_j N_k \delta_{jl}}{N^3 F_2(3,2) F_2(1,4)} 
  e^{-\tau_{\rm E} F_4(1,2,3,4)/\tau_{\rm D}},
  \label{eq:A4resb}  
\end{eqnarray}
where, again, we omitted terms proportional to $\sin r$.
Similarly, for $Q_4^{(\ref{fig:1b}b)}$ one finds
\begin{eqnarray}
  Q_4^{(\ref{fig:1b}b)} &=&
  \frac{N_i N_j N_l \delta_{ik}}{N^3 F_2(3,4) F_2(1,2)} 
  e^{-\tau_{\rm E} F_4(1,2,3,4)/\tau_{\rm D}}.
  \label{eq:A4resc}
\end{eqnarray}

Finally, if the encounter touches both lead openings, one has
\begin{eqnarray}
  \label{eq:A1}
  \label{eq:P40}
  Q_4^{(\ref{fig:0})} &=&
  \tau_{\rm D}
  \int dt_1 dt_2 \frac{P_{i}(t_1) P_{j}(t_2)}{t_{\rm enc}}
  \delta_{ik} \delta_{jl}
  \nonumber \\ && \mbox{} \times
  \int d(s_3-s_1) d(u_3-u_1)
  \frac{e^{i \Delta {\cal S}/\hbar}}{2 \pi \hbar},
\end{eqnarray}
where $P_i$ and $P_j$ are classical probabilities
given in Eq.\ (\ref{eq:Pij}), $t_1$ and $t_2$ are the propagation
times between the surface of section and the exit contact and the
entrance contact and the surface of section, respectively, and $t_{\rm
  enc} = t_1 + t_2$. The action difference $\Delta {\cal S}$ is given
by Eq.\ (\ref{eq:dS4}) with $t_{12} = t_{23} = t_{34} = t_{41} = 0$.
In order to ensure that $\alpha_1$ and $\alpha_3$ are correlated upon
entrance as well as exit, we require
\begin{equation}
  \label{eq:t12}
  t_1 < \frac{1}{\lambda} \ln \frac{c}{|s_3-s_1|},\ \
  t_2 < \frac{1}{\lambda} \ln \frac{c}{|u_3-u_1|}.
\end{equation}
Performing integrations to $t_1-t_2$ and $u_3-u_1$ and
 abbreviating $x = (s_3-s_1)c$, one finds
\begin{eqnarray}
  Q_4^{(\ref{fig:0})} &=&
  \delta_{ik} \delta_{jl} \frac{N_i N_j}{N^2}
  \int \frac{dt}{\tau_{\rm D}} e^{-t/\tau_{\rm D} + 
  i t(\varepsilon_1 - \varepsilon_2 + \varepsilon_3 - \varepsilon_4)}
  \nonumber \\ && \mbox{} \times
  \int_{0}^{e^{-\lambda t}}
  \frac{dx}{x} \frac{2 \sin(x r)}{\pi}.
\end{eqnarray}
The integral over $x$ has to be performed in the limit $r \to \infty$ 
at fixed ratio $\tau_{\rm E}/t$. Since
\begin{eqnarray}
  \int_0^{e^{-\lambda t}}  \frac{dx}{x} \frac{2 \sin(x r
  )}{\pi}
  &=& 
  \theta(\tau_{\rm E} - t)
\end{eqnarray}
if $r \to \infty$, 
where $\theta(x) = 1$ if $x > 0$ and $0$ otherwise, we find
\begin{eqnarray}
  \label{eq:A1res}
  Q_4^{(\ref{fig:0})} &=&
  \frac{N_i N_j \delta_{ik} \delta_{jl}}{N^2 F_4(1,2,3,4)}
  (1 - e^{-\tau_{\rm E}F_4(1,2,3,4)/\tau_{\rm D}}).
\end{eqnarray}
Together, Eqs.\ (\ref{eq:A4res}), (\ref{eq:A4resb}),
(\ref{eq:A4resc}), and (\ref{eq:A1res}) reproduce the prediction of
the effective random matrix theory.

We note that there is a close 
connection between the appearance of Kronecker deltas
involving the contact indices $i$, $j$, $k$, and $l$, and the
configurations of classical trajectories that contribute to $Q_4$: 
each encounter region that
touches a lead opening gives rise to a Kronecker delta and,
conversely, each Kronecker delta derived from an encounter that
touched a lead opening. Hence, the four contributions to $Q_4$ ---
the third line in Eq.\ (\ref{eq:Acl}) and the three terms in Eq.\
(\ref{eq:Armt2}) ---, each of which have a different product of
Kronecker deltas, are in a one-to-one correspondence with the four
configurations of classical trajectories shown in Figs.\
\ref{fig:1}, \ref{fig:1b}a, \ref{fig:1b}b, and \ref{fig:0},

We should point out that, without dependence on
the energy arguments, {\em i.e.}, after setting
$\varepsilon_1 = \varepsilon_2 = \varepsilon_3 = \varepsilon_4$, the
semiclassical calculation of $Q_4$ was first performed by Whitney and
Jacquod.\cite{kn:whitney2005} The same results 
can also be inferred from an earlier calculation by
Agam, Aleiner, and Larkin, who used a different
formalism.\cite{kn:agam2000}
The energy dependence of the four
terms  we calculate here is nontrivial, however, and clearly
reveals the structure of the classical
trajectories underlying the four different terms in the average.

While our calculation of 
$Q_4^{(\ref{fig:1})}$ closely followed that of Braun 
{\em et al.},\cite{kn:braun2005}
our calculation of the remaining three contributions to $Q_4$
differs considerably from that of Ref.\
\onlinecite{kn:braun2005}. Braun {\em et al.}\ do not consider 
encounters that touch the lead openings. Instead, they calculate
the remaining
contributions to $Q_4$ using the `diagonal 
approximation' for the trajectory sums. 
Although this approximation gives the
correct result if $\tau_{\rm E} \ll \tau_{\rm D}$, it is based on 
a different class of trajectories than the ones we considered (see 
also Ref.\ \onlinecite{kn:whitney2005}). In the diagonal
approximation, one considers the case that the four trajectories
$\alpha_1$, $\alpha_2$, $\alpha_3$, and $\alpha_4$ are pairwise equal
(rather than pairwise close),
but without imposing any relation between the two pairs. For example,
one admits trajectories $\alpha_1=\alpha_2$ and $\alpha_3=\alpha_4$
where $\alpha_2$ and $\alpha_3$ exit with the same transverse momentum
$p_{\perp}$ but at a classically different spatial coordinate $y$.
When summed over the full family of trajectories, such configurations
appear with rapidly oscillating 
phases, and their net contribution
vanishes.\cite{kn:rahav2006b,kn:whitney2005,kn:jacquod2006} 
These rapidly oscillating
phases disappear only if all four trajectories pass through the
contact at equal positions and angles (up to quantum uncertainties), 
as is the case for the configurations shown in Figs.\
\ref{fig:1b}a and b.

\subsection{Calculation of $Q_6$}

\begin{figure}
\epsfxsize=0.95\hsize
\epsffile{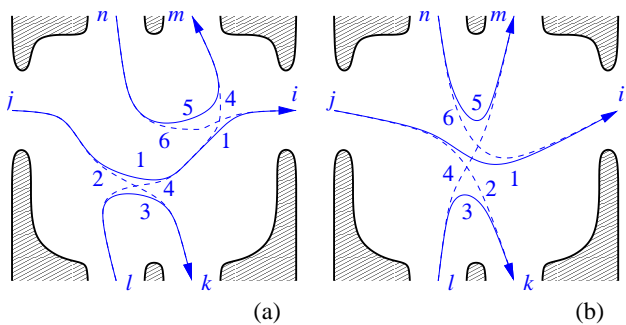}
\caption{\label{fig:3} Schematic drawing of six classical
  trajectories that contribute to $Q_6$ without encounters that
  touch the lead openings. The configuration shown in (a) consists of
  two separate two-encounters. There are two more
  configurations of this type, which are obtained by the cyclic 
  permutations $(1,2)  \to (3,4) \to (5,6)$. The configuration shown 
  in (b) consists of one three-encounter.}
\end{figure}

The generic configuration of classical
trajectories that contributes to $Q_6$ 
is shown in Fig.\ \ref{fig:3}.
Instead of dealing with trajectories with one small-angle
encounter, one now has to consider
configurations of trajectories with two small-angle encounters or of
trajectories with a
`three-encounter'.\cite{kn:mueller2004,kn:mueller2005,kn:heusler2006}
A three-encounter arises when
two encounters overlap, so that all six trajectories are within a
phase space distance $c$
simultaneously. In order to avoid confusion, we refer to non-overlapping
encounters of two pairs of trajectories as `two-encounters' for the 
remainder of this section. We first consider the case that all 
encounters fully reside inside the quantum dot. The case that
encounters touch the lead openings will be discussed afterwards.

Without encounters that touch the lead openings, one distinguishes two
types of contributions to $Q_6$: the contribution from the case
in which the trajectories undergo two
separate two-encounters, and the contribution from the case in which the
six classical trajectories involved undergo a three-encounter. These
are shown in Fig.\ \ref{fig:3}a and \ref{fig:3}b, respectively. The
configuration shown in Fig.\ \ref{fig:3}a shows only one
configuration with two two-encounters. There are two more
configurations, which are obtained by the cyclic permutations $(1,2,i,j) \to (3,4,k,l) \to (5,6,m,n)$.

Because the two two-encounters in Fig.\ \ref{fig:3}a do not overlap,
their contribution $Q_6^{(\ref{fig:3}a)}$ to $Q_6$ factorizes. The
stretch connecting the trajectories $\alpha_1$ and $\alpha_2$ to their
joint entrance contact $j$ contributes a factor $N_j/N F_2(1,2)$; the
other stretches connecting the encounters to the contacts contribute
similar factors to $Q_6^{(\ref{fig:3}a)}$. Further, using the results
of the previous subsection, we find that the two-encounter
of the trajectories $\alpha_1$, $\alpha_2$, $\alpha_3$, and
$\alpha_4$ in Fig.\ \ref{fig:3}a contributes a factor 
$-F_4(1,2,3,4) \exp(-\tau_{\rm E} F_4(1,2,3,4)/\tau_{\rm
  D})$. Similarly, the two-encounter involving the
trajectories $\alpha_1$, $\alpha_4$, $\alpha_5$, and $\alpha_6$ 
contributes a factor $-F_4(1,4,5,6) \exp(-\tau_{\rm E}
F_4(1,4,5,6)/\tau_{\rm D})$. Finally, the stretch of the trajectories
$\alpha_1$ and $\alpha_4$ that connects the two two-encounters in
Fig.\ \ref{fig:3}a contributes a factor $1/F_2(1,4)$. Using that 
$F_4(1,2,3,4) + F_4(1,4,5,6) = F_6(1,2,3,4,5,6) + F_2(1,4)$, we find
that $Q_6^{(\ref{fig:3}a)}$ is given by
%
%
%
\begin{widetext}
\begin{eqnarray}
  \label{eq:A3a}
  Q_6^{(\ref{fig:3}a)} &=&
  \frac{N_i N_j N_k N_l N_m N_ne^{-\tau_{\rm E} F_6(1,2,3,4,5,6)/\tau_{\rm D}}}{N^6 F_2(1,2) F_2(3,2) F_2(3,4)
  F_2(5,4) F_2(5,6) F_2(1,6)}
  \left[
  \frac{F_4(1,2,3,4) F_4(1,4,5,6)}{F_2(1,4)}
  e^{-\tau_{\rm E} F_4(1,4)/\tau_{\rm D}} + \ldots \right],
\end{eqnarray}
where the dots \ldots
represent two more terms obtained by the
cyclic permutations $(1,2,i,j) \to (3,4,k,l) \to (5,6,m,n)$. 
As before, the superscript
$(\ref{fig:3}a)$
refers to the figure that shows the corresponding configuration of
classical trajectories.

Although the calculation of $Q_6^{(\ref{fig:3}b)}$ is similar in
spirit, it proves
rather cumbersome. Referring to App.\ \ref{app:B} for details, we find
\begin{eqnarray}
  Q_6^{(\ref{fig:3}b)} &=&
  \frac{N_i N_j N_k N_l N_m N_n
  e^{-\tau_{\rm E} F_6(1,2,3,4,5,6)/\tau_{\rm D}}
  }{N^6 F_2(1,2) F_2(3,2) F_2(3,4)
  F_2(5,4) F_2(5,6) F_2(1,6)}
  \left\{
  \frac{F_4(1,2,3,4) F_4(1,4,5,6)}{F_2(1,4)}
  \left[1 - e^{-\tau_{\rm E}F_2(1,4)/\tau_{\rm D}} \right] + \ldots \right\}
  \nonumber \\ && \mbox{}
  - \frac{N_i N_j N_k N_l N_m N_n e^{-\tau_{\rm
  E}F_6(1,2,3,4,5,6)/\tau_{\rm D}}}
  {N^6 F_2(1,2) F_2(3,2) F_2(3,4)
  F_2(5,4) F_2(5,6) F_2(1,6)}F_6(1,2,3,4,5,6).  
\end{eqnarray}
Combining these two contributions, we find that the terms proportional
to $\exp(-2 \tau_{\rm E}/\tau_{\rm D})$ cancel, so that the final
result is uniformly proportional to $\exp(-\tau_{\rm E}/\tau_{\rm D})$,
\begin{eqnarray}
  \label{eq:P62}
  Q_6^{(\ref{fig:3})} &=&
  \frac{N_i N_j N_k N_l N_m N_n
  e^{-\tau_{\rm E} F_6(1,2,3,4,5,6)/\tau_{\rm D}}}{N^6
  F_2(1,2) F_2(3,4) F_2(5,6)
  F_2(1,6) F_2(3,2) F_2(5,4)}
  \nonumber \\ && \mbox{} \times
  \left[
  \frac{F_4(1,2,3,4) F_4(1,6,5,4)}{F_2(1,4)} +
  \frac{F_4(3,4,5,6) F_4(3,2,1,6)}{F_2(3,6)} +
  \frac{F_4(5,6,1,2) F_4(5,4,3,2)}{F_2(5,2)}
  + F_6(1,2,3,4,5,6) \right].
  \nonumber \\ 
\end{eqnarray}

\begin{figure}
\epsfxsize=0.95\hsize
\epsffile{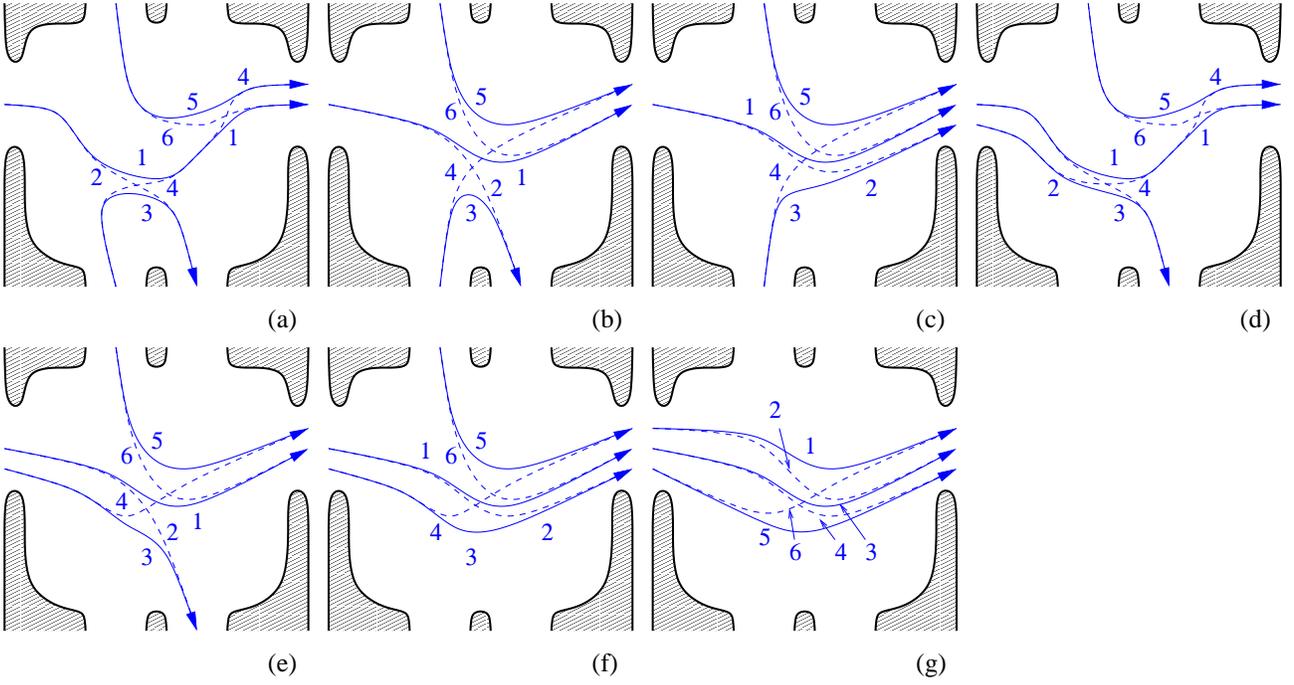}
\caption{\label{fig:4}
  \label{fig:7}
  \label{fig:8}
  \label{fig:10}
Schematic drawing of six classical
trajectories that provide the remaining contributions to $Q_6$. 
The configuration shown in (a) consists of
two separate two-encounters, where one two-encounter touches a lead
opening. The configuration shown in (b) consists of one
three-encounter, which partially touches one lead opening. The
configuration of (c) consists of a three-encounter that fully touches
one of the lead openings. The configuration of (d) has two
two-encounters that each touch lead openings. The configuration of (e)
has a three-encounter that partially touches two lead openings. The
configuration of (f) has a three-encounter that partially touches one,
and fully touches the other lead opening. Finally, the configuration
of (g) fully touches both lead openings. The
contributions from (a) and (b) have one Kronecker delta involving the
lead indices, the contributions from (c), (d), and (e) have two
Kronecker deltas, the contribution of (f) is zero, and the
contribution of (g) has four Kronecker deltas.}
\end{figure}

In order to compute the contribution to $Q_6$ of encounters that touch
the lead openings, we have to consider the configurations of classical
trajectories shown in Fig.\ \ref{fig:4}. Figure \ref{fig:4}a 
show trajectories with two two-encounters where one encounter touches the
lead opening. Figure \ref{fig:4}b shows a configuration of 
classical trajectories with a three-encounter where the
three-encounter partially touches a lead opening. In both cases, the
escape of only one pair of trajectories is correlated, so that the
contribution of the configurations of Figs.\ \ref{fig:4}a and b has
one Kronecker delta. In addition to the configurations shown in
Fig.\ \ref{fig:4}a and b, there five more contributions to $Q_6$ that 
are obtained by cyclic permutations of the trajectory labels and by the
exchange of entrance and exit contacts. 
As before, the
contribution from the configuration with two two-encounters factorizes,
\begin{eqnarray}
  Q_6^{(\ref{fig:4}a)} &=&
  - \frac{N_j N_l N_n e^{-\tau_{\rm E} F_6(1,2,3,4,5,6)/\tau_{\rm D}}}
  {N^3 F_2(1,2) F_2(3,4)
  F_2(5,6)}
  \left[ \frac{N_i N_k F_4(1,2,3,4) \delta_{im}}{N^2 F_2(2,3)
  F_2(1,4)}
  e^{-\tau_{\rm E} F_2(1,4)/\tau_{\rm D}} + \ldots \right]
  \nonumber \\ && \mbox{} -
  \frac{N_i N_k N_m
  e^{-\tau_{\rm E}F_6(1,2,3,4,5,6)/\tau_{\rm D}}}
  {N^3 F_2(3,2) F_2(5,4)
  F_2(1,6)}
  \left[ \frac{N_j N_l F_4(3,4,5,2) \delta_{jn}}{N^2
  F_2(3,4) F_2(5,2)} e^{-\tau_{\rm E} F_2(5,2)/\tau_{\rm D}}
  + \ldots \right],
\end{eqnarray}
Details of the calculation of the three-encounter are again
left to the appendix. The result is
\begin{eqnarray}
  Q_6^{(\ref{fig:4}b)} &=&
  \label{eq:P6b}
  -\frac{N_j N_l N_n 
  e^{-\tau_{\rm E} F_6(1,2,3,4,5,6)/\tau_{\rm D}}}
  {N^3 F_2(1,2) F_2(3,4)
  F_2(5,6)}
  \left[ \frac{N_i N_k F_4(1,2,3,4) \delta_{im}}{N^2 F_2(2,3)
  F_2(1,4)}
  \left(1-e^{-\tau_{\rm E} F_2(1,4)/\tau_{\rm D}}\right) + \ldots \right]
  \nonumber \\ && \mbox{} -
  \frac{N_i N_k N_m
  e^{-\tau_{\rm E}F_6(1,2,3,4,5,6)/\tau_{\rm D}}}
  {N^3 F_2(3,2) F_2(5,4)
  F_2(1,6)}
  \left[ \frac{N_j N_l F_4(3,4,5,2) \delta_{jn}}{N^2
  F_2(3,4) F_2(5,2)} \left(1-e^{-\tau_{\rm E} F_2(5,2)/\tau_{\rm D}}
  \right)
  + \ldots \right].
\end{eqnarray}
Combining both contributions, we again find that the terms
proportional to $\exp(-2\tau_{\rm E}/\tau_{\rm D})$ cancel, so that
\begin{eqnarray}
  \label{eq:P63}
  Q_6^{(\ref{fig:4}ab)} &=&
  \left[\frac{N_i N_j N_k N_l N_n \delta_{im} F_4(1,2,3,4)}
  {N^5 F_2(1,2) F_2(3,2) F_2(3,4) F_2(5,6) F_2(1,4)}
  +
  \frac{N_i N_j N_k N_l N_m \delta_{jn}F_4(3,4,5,2)}
  {N^5 F_2(3,2) F_2(3,4) F_2(5,4)
  F_2(1,6) F_2(5,2)}
  + \ldots \right]
  \nonumber \\ && \mbox{} \times
  e^{-\tau_{\rm E}F_6(1,2,3,4,5,6)/\tau_{\rm D}}.
\end{eqnarray}

There are two different types of contributions to $Q_6$ with two
Kronecker deltas: Two Kronecker deltas that both involve lead indices
of entrance (or exit)
leads, or a product of two Kronecker deltas, where one involves
lead indices of the entrance lead and one involves lead indices of the
exit lead. Only a configuration of trajectories with a three-encounter
contribute to the former, see Fig.\ \ref{fig:7}c, whereas
configurations of trajectories with a three-encounter as well as
configurations of trajectories with two
two-encounters contribute to the latter, see Fig.\
\ref{fig:8}d and e. Referring to the appendix for calculational details
regarding the configurations of trajectories with a three-encounter,
here we simply quote the results,
\begin{eqnarray}
  \label{eq:P67}
  Q_6^{(\ref{fig:7}c)} &=&
  \left[
  \frac{N_i N_j N_l N_n \delta_{ik} \delta_{im}}{N^4 F_2(1,2) F_2(3,4)
  F_2(5,6)} +
  \frac{N_i N_j N_k N_m \delta_{jl} \delta_{jn}}{N^4 F_2(3,2) F_2(5,4)
  F_2(1,6)} \right] e^{-\tau_{\rm E} F_6(1,2,3,4,5,6)/\tau_{\rm D}},
  \\
  \label{eq:P68}
  Q_6^{(\ref{fig:8}de)} &=&
  \left[
  \frac{N_i N_j N_l N_m \delta_{ik} \delta_{ln}}{N^4 F_2(1,2) F_2(3,6)
  F_2(5,4)} +
  \frac{N_i N_j N_k N_n \delta_{im} \delta_{jl}}{N^4 F_2(1,4) F_2(3,2)
  F_2(5,6)} +
  \frac{N_i N_j N_k N_l \delta_{jn} \delta_{km}}{N^4 F_2(1,6) F_2(3,4)
  F_2(5,2)}  \right]
  \nonumber \\ && \mbox{} \times
   e^{-\tau_{\rm E} F_6(1,2,3,4,5,6)/\tau_{\rm D}}.
\end{eqnarray}
\end{widetext}


Finally, we have to consider the case that there is correlated escape
of one pair and one triplet of trajectories
and correlated escape of two triplets of trajectories, see Figs.\
\ref{fig:10}f and g, respectively.
One verifies that there is no contribution to $Q_6$ for the
configuration shown in Fig.\ \ref{fig:10}f. For the configuration
shown in Fig.\ \ref{fig:10}g one finds
\begin{eqnarray}
  \label{eq:P61}
  Q_6^{(\ref{fig:10}g)} &=&
  \frac{N_i N_j \delta_{ik} \delta_{im} \delta_{jl} \delta_{jn}}{N^2
  F_6(1,2,3,4,5,6)}
  (1 - e^{-\tau_{\rm E} F_6(1,2,3,4,5,6)/\tau_{\rm D}}).
  \nonumber \\
\end{eqnarray}
Together, Eqs.\ (\ref{eq:P62}), (\ref{eq:P63}), (\ref{eq:P67}),
(\ref{eq:P68}), and (\ref{eq:P61}) combine to the effective random
matrix theory ansatz for $Q_6$.

\section{Conclusion}
\label{sec:4}

In this article we calculated the Ehrenfest-time dependence of the
ensemble averages of polynomial functions $Q_n$ of the $N \times N$
scattering matrix $S$
of a ballistic chaotic quantum dot and its hermitian conjugate
$S^{\dagger}$, for polynomials of degree $n=2$, $4$, and $6$. For all
cases our calculations of the leading-in-$N$ averages
agreed with the effective random matrix theory
of Silvestrov, Goorden, and Beenakker.\cite{kn:silvestrov2003b} 

The detailed semiclassical calculations of $Q_2$ and $Q_4$ were
included in this article because they are a prerequisite for the 
semiclassical calculations of $Q_6$. For the calculation of $Q_4$ 
a few guiding principles would have been sufficient, however. These
are: (i) the classical trajectories contributing to 
$Q_4$ have at most one small-angle encounter, (ii) each encounter
lasts one Ehrenfest time $\tau_{\rm E}$, and
(iii) $Q_4$ becomes equal to the classical limit $Q_4^{\rm cl}$ if 
$\tau_{\rm E}$ is much larger than the mean dwell time $\tau_{\rm D}$,
whereas $Q_4$ equals the random matrix limit $Q_4^{\rm RMT}$ if
$\tau_{\rm E} \ll \tau_{\rm D}$. From the first two
guiding principles, both of which are well established in the
literature,\cite{kn:aleiner1996,kn:agam2000} and the exponential
distribution of classical dwell times, one concludes that 
$Q_4$ is of the form 
\begin{equation}
  Q_4 = A + B \exp(-\tau_{\rm E}/\tau_{\rm
D}).
\end{equation}
The third guiding principle then fully determines the
 coefficients $A$ and $B$ and, hence, $Q_4$.

No such shortcut exists for the calculation of $Q_6$. The relevant
configurations of semiclassical trajectories have up to two
encounters, hence we expect a general form 
\begin{equation}
  Q_6 = A + B 
\exp(-\tau_{\rm E}/\tau_{\rm D}) + C \exp(-2 \tau_{\rm E}/\tau_{\rm
D}).
  \label{eq:ABC}
\end{equation}
The detailed calculation of this article was needed to show that
the coefficient $C = 0$. In this sense, the semiclassical calculation
of $Q_6$ provides the first nontrivial test of the effective random matrix 
theory.

The fact that the coefficient $C$ in Eq.\ (\ref{eq:ABC}) is zero is
at the heart of the effective random matrix theory ansatz: It implies
that (i) all trajectories with dwell times longer than $\tau_{\rm E}$ are
treated equally and (ii) that $\tau_{\rm E}$ is the only time marking
the boundary between classical and quantum effects. Whereas our 
calculation indirectly verifies these claims for
leading-in-$N$ averages of low-degree
polynomials of $S$ and $S^{\dagger}$, they are not true for subleading-in-$N$
averages. For subleading-in-$N$ averages, more than one integer
multiple of $\tau_{\rm E}$ may determine quantum effects. Known
examples are weak localization, to which only trajectories with a 
dwell time larger than $2 \tau_{\rm E}$ 
contribute, although the weak localization correction is suppressed 
$\propto \exp(-\tau_{\rm E}/\tau_{\rm 
D}$) with an exponent containing $\tau_{\rm E}$, not $2 \tau_{\rm 
E}$,\cite{kn:aleiner1996,kn:richter2002,kn:rahav2005}
the leading quantum correction to the spectral form factor $K(\omega)$
of a quantum dot with broken time-reversal
symmetry,\cite{kn:tian2004b} which has oscillations
proportional to $\exp(2i \omega \tau_{\rm E})$, $\exp(3 i \omega
\tau_{\rm E})$, and $\exp(4 i \omega \tau_{\rm E})$, 
conductance fluctuations in a bulk chaotic conductor (rather than a
chaotic quantum dot),\cite{kn:tian2006} and the mean square current pumped
through a chaotic quantum dot with time-dependent 
potentials.\cite{kn:rahav2006c}

Having verified the effective random matrix theory ansatz for
leading-in-$N$ averages for polynomials in $S$ and $S^{\dagger}$ up to
degree six, one wonders whether this verification can be extended to
polynomials of arbitrary degree. 
In our
calculation, the simple result predicted by the effective random
matrix theory
was obtained only after a tedious calculation showing that the
coefficient $C$ in Eq.\ (\ref{eq:ABC}) above is zero.
The possibility of the appearance of a term $\propto
\exp(-2\tau_{\rm E}/\tau_{\rm D})$ caused a large increase
in complexity upon going from polynomials of degree four to
polynomials of degree six.  We have searched for a
simplification in our calculation that would allow us to cancel all
contributions $\propto \exp(-2\tau_{\rm E}/\tau_{\rm D})$
at an earlier point in the calculation. However, since
the terms proportional to $\exp(-2 \tau_{\rm E}/\tau_{\rm D})$ were
obtained from qualitatively different configurations of classical
trajectories, we have not been able to find such a simplification.
Without such a simplification, the task of extending the present
calculation to polynomials of arbitrary degree seems too daunting to
accomplish.


\acknowledgments

We thank Maxim Vavilov for discussions.
This work was supported by the NSF under grant no.\ DMR 0334499 and 
by the Packard Foundation.

\appendix

\section{Trajectory sum with fixed coordinates $s$ and $u'$}
\label{app:A}

In this appendix we show that one can replace the standard formulation
in which the traces $Q_n$ are expressed in terms of integrals over
classical trajectories with specified transverse momentum components
at entrance and exit contacts by an integral over classical
trajectories with specified stable and unstable phase space
coordinates at entrance and exit contacts, respectively. 

We first consider a single integral over the transverse momentum
$p_{\perp}$ in the entrance contact,
\begin{equation}
  K(p_{\perp,1}',p_{\perp,2}') = \frac{1}{2 \pi \hbar}
  \int dp_{\perp} \sum_{\alpha_1,\alpha_2}
  A_1 A_2 e^{i ({\cal S}_1 - {\cal S}_2)/\hbar},
  \label{eq:K}
\end{equation}
where the trajectory $\alpha_{\mu}$ has transverse momentum
$p_{\perp}$ at its entrance contact,
\begin{equation}
  p_{\perp,1} = p_{\perp,2} = p_{\perp},
  \label{eq:app:pcond}
\end{equation}
and transverse momentum
$p_{\perp,\mu}'$ at its exit contact, $\mu=1,2$. The stability
amplitude is given by Eq.\ (\ref{eq:Aampl}). The classical action
${\cal S}_{\mu}$ is a function of $p_{\perp}$ and $p_{\perp}'$,
$\mu=1,2$.

\begin{figure}
\epsfxsize=0.45\hsize
\hspace{0.05\hsize}
\epsffile{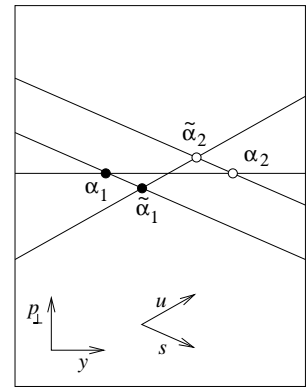}
\caption{\label{fig:0p} Poincar\'e surface of section of lead opening,
  together with the standard coordinates $y$, $p_{\perp}$, as well as
  the phase space coordinates $s$, $u$, taken along the stable and
  unstable directions in phase space. The trajectory pair $\alpha_1$,
  $\alpha_2$, which have equal perpendicular momentum components
  $p_{\perp,1} = p_{\perp,2}$ are mapped to a trajectory pair
  $\tilde \alpha_1$, $\tilde \alpha_2$ 
  with equal stable phase space coordinates,
  $\tilde s_1 = \tilde s_2$. 
  The unstable phase space coordinates do not change in
  the mapping.}
\end{figure}

In order to rewrite $K$ in terms of an integral over classical
trajectories with a fixed stable phase space coordinate $s$ upon
entrance into the quantum dot, we consider a Poincar\'e surface of 
section taken at the entrance contact, see Fig.\ \ref{fig:0p}. There 
are two
sets of coordinates to label trajectories piercing this Poincar\'e
surface of section: the spatial and momentum coordinates $y$ and
$p_{\perp}$, and the stable and unstable phase space coordinates $s$
and $u$, see Fig.\ \ref{fig:0p}. The coordinate axes for $s$ and $u$
do not need to be perpendicular, but the coordinates are normalized
such that $ds du = dy dp_{\perp}$.\cite{kn:spehner2003} 
The phase space points where
$\alpha_1$ and $\alpha_2$ pierce the Poincar\'e surface of section are
indicated in the figure.

To each pair of two sufficiently close
classical trajectories  $\alpha_1$, $\alpha_2$ we now assign another pair 
$\tilde \alpha_1$, $\tilde \alpha_2$, where the trajectory
$\tilde \alpha_\mu$ is obtained from $\alpha_{\mu}$ by 
slightly changing the boundary condition in the entrance contact while
keeping the boundary condition upon exit the same. This implies that 
the unstable phase space coordinates of $\alpha_\mu$ and $\tilde
\alpha_\mu$ are equal, $\mu=1,2$. In order to uniquely define the pair
$\tilde \alpha_1, \tilde \alpha_2$, we require that $\tilde \alpha_1$
and $\tilde \alpha_2$ have the same stable phase space coordinate $s$ 
upon entry, and that $p_{\perp,\tilde 1} - p_{\perp,1} = p_{\perp,2} -
p_{\perp,\tilde 2}$. The construction of
$\tilde \alpha_1$ and $\tilde \alpha_2$ is shown in Fig.\ \ref{fig:0p}.

For this construction, one verifies that the action difference 
of the original and 
primed trajectory pair is not changed,
\begin{equation}
  S_{1} - S_{2} =
  \tilde S_{\tilde 1} - \tilde S_{\tilde 2},
\end{equation}
provided we replace the action ${\cal S}$, which is a function of
$p_{\perp}$, by the action $\tilde S$, which is a function of the 
stable phase space coordinate $s$, obtained by Legendre transform of 
the original action ${\cal S}$. 
Also, using the normalization $ds du = dp_{\perp} dy$ and Eq.\
(\ref{eq:app:pcond}), one finds
\begin{eqnarray}
  \lefteqn{
  \int dp_{\perp} 
  \left| \frac{\partial p_{\perp,1}'}{\partial y_1} 
  \right|^{-1/2}
  \left| \frac{\partial p_{\perp,2}'}{\partial y_2} 
  \right|^{-1/2} \ldots =} \nonumber \\ && 
  \int ds   \left| \frac{\partial p_{\perp,\tilde 1}'}{\partial
  u_{\tilde 1}} 
  \right|^{-1/2}
  \left| \frac{\partial p_{\perp,\tilde 2}'}{\partial u_{\tilde 2}} 
  \right|^{-1/2} \ldots,
\end{eqnarray}
where
\begin{equation}
  s = s_{\tilde 1} = s_{\tilde 2}.
\end{equation}

The same procedure can be applied to integrations over the transverse
momentum $p_{\perp}'$ in the exit contacts. In this case, one replaces
the integral over $p_{\perp}'$ by an integral over the unstable phase
space coordinate $u'$, and replaces pairs of trajectories with equal
$p_{\perp}'$ by pairs of trajectories with equal $u'$. The result of
this program is precisely Eqs.\ (\ref{eq:Qn2})--(\ref{eq:Qn6})
of the main text.
  
\section{Three-encounter of classical trajectories}
\label{app:B}

\begin{figure}
\epsfxsize=0.6\hsize
\hspace{0.05\hsize}
\epsffile{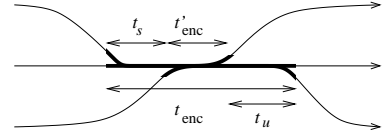}
\caption{\label{fig:enctimes}
  Definition of the four measures of the encounter duration
  used in the text: $t_{\rm enc}$ is the total duration of the
  three-encounter, $t_{\rm enc}'$ is the duration of the part of the
  encounter that all trajectories involved in the three-encounter are
  within a phase-space distance $c$, and $t_s$ and $t_u$ are the
  durations of the encounter stretches where one pair of trajectories
  has already diverged from the remaining two pairs.}
\end{figure}

\subsection{Three-encounter fully inside the quantum dot}

In this appendix we describe the details of the calculation of
$Q_6^{(\ref{fig:3}b)}$. This is the contribution to $Q_6$
from a three-encounter that fully resides inside the quantum dot.

In order to calculate $Q_6^{(\ref{fig:3}b)}$, we take
a Poincar\'e surface of section at a point that all six
trajectories are within a phase space distance $c$. We parameterize
the trajectories using phase space coordinates $s$ and $u$
along the stable and unstable directions in phase space.
At the Poincar\'e surface of section, one has
\begin{eqnarray}
  s_1 &=& s_2, \ \ s_3 = s_4, \ \ s_5 = s_6 \nonumber \\
  u_2 &=& u_3, \ \ u_4 = u_5, \ \ u_6 = u_1.
\end{eqnarray}
(These equalities continue to hold if the three-encounter touches the
 lead opening.) The action difference $\Delta {\cal S}$
 reads\cite{kn:mueller2004,kn:mueller2005}
\begin{eqnarray}
  \label{eq:dS3}
  \Delta {\cal S} &=& (u_3-u_1)(s_3-s_1) + (u_5-u_1)(s_5-s_3) 
  \nonumber \\ && \mbox{}
  + 
  \varepsilon_1 t_{1}
  - \varepsilon_2 t_{2} + \ldots - \varepsilon_6 t_{6}.
\end{eqnarray}
In order to parameterize the durations of the six
trajectories involved, we separate each duration $t_{\mu}$,
$\mu=1,2,\ldots,6$, into segments before and after the encounter and one
segment of duration
$t_{\mu,{\rm enc}}$ inside the encounter region. Similarly, we
define
\begin{eqnarray}
  \label{eq:Sencdef}
  \Delta {\cal S}_{\rm enc} &=& (u_3-u_1)(s_3-s_1) + (u_5-u_1)(s_5-s_3) 
  \nonumber \\ && \mbox{}
  + 
  \varepsilon_1 t_{1,{\rm enc}}
  - \varepsilon_2 t_{2,{\rm enc}} + \ldots - \varepsilon_6 t_{6,{\rm enc}}.
\end{eqnarray}
Integrating over the time segments outside the encounter, we then 
arrive at
\begin{widetext}
\begin{eqnarray}
  Q_6^{(\ref{fig:3}b)} &=&
  \frac{N_i N_j N_k N_l N_m N_n}{N^6
  F_2(1,2) F_2(3,4) F_2(5,6)
  F_2(1,6) F_2(3,2) F_2(5,4)} I,
  \label{eq:A3b}
\end{eqnarray}
with
\begin{eqnarray}
  I &=&
  \tau_{\rm D} \int \frac{d(s_3-s_1) d(s_5-s_1) d(u_3-u_1) d(u_5-u_1)}{(2 \pi \hbar)^2
  t_{\rm enc}'}
  e^{i \Delta {\cal S}_{\rm enc}/\hbar - t_{\rm enc}/\tau_{\rm D}}.
  \label{eq:A3}
\end{eqnarray}
Here $t_{\rm enc}$ is the total duration of the encounter (the time
that at least two trajectories are close together),
\begin{eqnarray}
  t_{\rm enc} &=&
  \frac{1}{\lambda} \ln \frac{c}{\min(|s_3-s_1|,|s_5-s_1|,|s_3-s_5|)}
  + 
  \frac{1}{\lambda} \ln \frac{c}{\min(|u_3-u_1|,|u_5-u_1|,|u_3-u_5|)},
\end{eqnarray}
and $t_{\rm enc}'$ is the time that all three trajectories are close
together, see Fig.\ \ref{fig:enctimes},
\begin{eqnarray}
  t_{\rm enc}'  &=&
  \frac{1}{\lambda} \ln \frac{c}{\max(|s_3-s_1|,|s_5-s_1|,|s_3-s_5|)}
  + 
  \frac{1}{\lambda} \ln \frac{c}{\max(|u_3-u_1|,|u_5-u_1|,|u_3-u_5|)}.
\end{eqnarray}
The factor $t_{\rm enc}'$ in the denominator cancels a spurious
contribution to the integral arising from the freedom to choose the
Poincar\'e surface of section at any point during three-encounter. 
The integration domain in Eq.\ (\ref{eq:A3}) is given by the conditions
$\max(|u_3-u_1|,|u_5-u_1|,|u_3-u_5|) < c$ and
$\max(|s_3-s_1|,|s_5-s_1|,|s_3-s_5|) < c$. 

We rewrite the integral (\ref{eq:A3}) using
$s =
\max(|s_3-s_1|,|s_5-s_1|,|s_3-s_5|)$, $s' =
\min(|s_3-s_1|,|s_5-s_1|,|s_3-s_5|$, $u =
\max(|u_3-u_1|,|u_5-u_1|,|u_3-u_5|)$ and $u' =
\min(|u_3-u_1|,|u_5-u_1|,|u_3-u_5|)$ as our integration
variables. After some tedious algebra, one arrives at
\begin{eqnarray}
  I &=& \tau_{\rm D}
  \int_0^{c} ds du \int_0^{s/2} ds' \int_0^{u/2} du'
  \frac{e^{-t_{\rm enc}'F_6(1,2,3,4,5,6)/\tau_{\rm D}}}{(\pi \hbar)^2
  t_{\rm enc}'} 
  \nonumber \\ && \mbox{} \times
  \left\{  
  \left[ e^{-F_4(1,2,5,6) t_s/\tau_{\rm D} -F_4(1,6,3,2) t_u/\tau_{\rm
  D}}
  +
  e^{-F_4(1,2,3,4) t_s/\tau_{\rm D} - F_4(1,6,3,2) t_u/\tau_{\rm D}}
  + \ldots
  \right]
  \right. \nonumber \\ && \ \ \ \ \left. \mbox{} \times
  \left( \cos\frac{us' + u's - us}{\hbar}
  + \cos\frac{u s + u' s' - u's}{\hbar} +
  \cos\frac{u s + u' s' - u s'}{\hbar} +
  \cos\frac{u s - u' s'}{\hbar} \right) 
  \right. \nonumber \\ &&  \ \ \left. \mbox{}
  + 
  2
  \left[ e^{-F_4(1,2,3,4) t_s/\tau_{\rm D} - F_4(5,4,1,6)
  t_u/\tau_{\rm D}}
  + \ldots
  \right]
  \left( \cos\frac{u s' - u' s' + s u'}{\hbar} +
  \cos\frac{s' u - s u'}{\hbar} \right) \right\},
\end{eqnarray} 
\end{widetext}
where the dots indicate cyclic permutations $(1,2) \to (3,4) \to
 (5,6)$, and with
\begin{eqnarray}
  t_{\rm enc}' &=& \frac{1}{\lambda} \ln \frac{c}{u} +
  \frac{1}{\lambda} \ln \frac{c}{s}, \nonumber \\
  t_{s} &=& \frac{1}{\lambda} \ln \frac{s}{s'}, \nonumber \\
  t_{u} &=& \frac{1}{\lambda} \ln \frac{u}{u'}.
\end{eqnarray}
The total duration of the encounter is $t_{\rm enc} = t_{\rm enc}' +
t_s + t_u$.
We then perform the variable change
\begin{eqnarray}
  s &=& c/\sigma,\ \ s' = c x'/\sigma, \nonumber \\
  u &=& c y \sigma,\ \ u' = c y y' \sigma.
\end{eqnarray}
With these new variables, the integration over $\sigma$ can be done
and cancels the factor $t_{\rm enc}'$ in the denominator. 
Writing $r =
c^2/\hbar$, the remaining integral then reads
\begin{widetext}
\begin{eqnarray}
  I &=& \frac{\lambda \tau_{\rm D} r^2}{\pi^2}
  \int_0^{1/2} dx' dy' \int_0^{1} y dy
  y^{F_6(1,2,3,4,5,6)/\lambda \tau_{\rm D}}
  \nonumber \\ && \mbox{} \times
  \left\{
  \left[(x')^{F_4(1,2,5,6)/\lambda \tau_{\rm D}}
  (y')^{F_4(1,6,3,2)/\lambda \tau_{\rm D}}
  + (x')^{F_4(1,2,3,4)/\lambda \tau_{\rm D}}
  (y')^{F_4(1,6,3,2)/\lambda \tau_{\rm D}} 
  + \ldots
  \right]
  \right. \nonumber \\ && \left. \mbox{} \times
  \left[ \cos(y r(x' + y' - 1)) + \cos(y r(1 + x' y' - y'))
  + \cos(y r(1 + x' y' - x')) + \cos(y r(1 - x' y')) \right]
  \right. \nonumber \\ && \left. \mbox{} +
  2 \left[ 
  (x')^{F_4(1,2,3,4)/\lambda \tau_{\rm D}}
  (y')^{F_4(5,4,1,6)/\lambda \tau_{\rm D}} + \ldots
  \right]
  \left[ \cos(y r(x' + y' - x' y') + \cos(y r(x' - y')) \right]
  \right\}.
  \label{eq:F2six}
\end{eqnarray}

In order to evaluate Eq.\ (\ref{eq:F2six}), we write each product
$(x')^{\epsilon_1} (y')^{\epsilon_2}$ as
\begin{eqnarray}
  (x')^{\epsilon_1} (y')^{\epsilon_2} &=&
  (1/2)^{\epsilon_1+\epsilon_2}
  +
  [(x')^{\epsilon_1} - (1/2)^{\epsilon_1}](1/2)^{\epsilon_2}
  +
  [(y')^{\epsilon_2} - (1/2)^{\epsilon_2}](1/2)^{\epsilon_1}
  +
  [(x')^{\epsilon_1} - (1/2)^{\epsilon_1}][(y')^{\epsilon_2} -
  (1/2)^{\epsilon_2}]
  \nonumber \\ 
  \label{eq:su}
\end{eqnarray}
and evaluate each of the four terms in Eq.\ (\ref{eq:su})
separately. In Eq.\ (\ref{eq:su}), the exponents $\epsilon_1$ and
$\epsilon_2$ represent the exponents $F_4(1,2,5,6)/\lambda \tau_{\rm
  D}$ etc. in Eq.\ (\ref{eq:F2six}). The separation in Eq.\
(\ref{eq:su}) makes sense, because both $\epsilon_1$ and $\epsilon_2$
are sent to zero if the classical limit $\hbar \to 0$ at fixed ratio
$\tau_{\rm E}/\tau_{\rm D}$ is taken. Hence, the first term in Eq.\
(\ref{eq:su}) approaches unity, whereas the other terms are nonzero
only if $x'$ or $y'$ are small, or both. 

The four terms in Eq.\ (\ref{eq:su}) are evaluated separately.
For the first term in Eq.\ (\ref{eq:su}), the integrals to $x'$ and
$y'$ can be done exactly. Omitting a prefactor $(1/2)^{\epsilon_1 +
  \epsilon_2}$, which is sent to $1$ in the classical limit taken
here, and abbreviating $\eta = F_6(1,2,3,4,5,6)/\lambda \tau_{\rm D}$,
we find that the integrals from the first term give
\begin{eqnarray}
  I_1 &=&
  \frac{6 \lambda \tau_{\rm D}r^2}{\pi^2}
  \int_0^{1/2} dx' dy' \int_0^1 dy y y^{\eta}
  [\cos(y r(x' + y' - 1)) + \cos(y r(1 + x' y' - y'))
  \nonumber \\ && \mbox{}
  + \cos(y r(1 + x' y' - x')) + \cos(y r(1 - x' y'))
  + \cos(y r(x' + y' - x' y')) + \cos(y r(x' - y'))] \nonumber \\
  &=&
    \frac{6 \lambda \tau_{\rm D}}{\pi^2}
  \int_0^{1} dy  y^{\eta}
  \frac{\partial}{\partial y}
  \left[-g_c(r y) \cos(r y) + g_s(r y) \sin(r y) \right],
  \nonumber \\
\end{eqnarray}
where
\begin{equation}
  g_c(v) = \int_0^v dw \frac{1 - \cos w}{w}, \ \
  g_s(v) = \int_0^v dw \frac{\sin w}{w}.
\end{equation}
Then, performing a partial integration to $y$ and omitting terms that
are oscillating with $r$, we find  
\begin{eqnarray}
  I_1
  &=&
  \frac{6 \lambda \tau_{\rm D} \eta}{\pi^2} 
  \int_0^1 \frac{dy}{y} y^{\eta}
  \left[g_c(r y) \cos(r y) - g_s(r y) \sin(r y) \right],
\end{eqnarray}
In this expression we can take the limits $r \to \infty$ and $\eta \to
0$, keeping $r^{-\eta} = e^{-\tau_{\rm E} F_6(1,2,3,4,5,6)/\tau_{\rm D}}$
fixed. We then find
\begin{eqnarray}
  I_1
  &=&
  - F_6(1,2,3,4,5,6)
  e^{-\tau_{\rm E} F_6(1,2,3,4,5,6)/\tau_{\rm D}}.
\end{eqnarray}
 
For the second term in Eq.\ (\ref{eq:su}),
we first perform the integration to $y'$, with the result
\begin{eqnarray}
  \label{eq:I21}
  I_2 &=&
  \frac{\lambda \tau_{\rm D} r^2}{\pi^2}
  \int_0^{1/2} dx' dy' \int_0^1 dy y y^{\eta} [(x')^{\epsilon}-(1/2)^{\epsilon}]
  [\cos(y r(x' + y' - 1)) + \cos(y r(1 + x' y' - y'))
  \nonumber \\ && \mbox{}
  + \cos(y r(1 + x' y' - x')) + \cos(y r(1 - x' y'))
  + \cos(y r(x' + y' - x' y')) + \cos(y r(x' - y'))] 
  + \ldots \nonumber \\
  &=&
  \frac{\lambda \tau_{\rm D}}{\pi^2}
  \int_0^{1/2} dx' \int_0^{1} dy y^{\eta} [(x')^{\epsilon}-(1/2)^{\epsilon}]
  \frac{\partial}{\partial y}
  \frac{x' \cos(y r x') - \cos(r y) + (1- x') \cos(y
  r(1-x'))}{x'(1-x')} + \ldots,
\end{eqnarray}
where we abbreviated $\epsilon = F_4(1,2,3,4)/\lambda \tau_{\rm D}$.
Performing a partial integration to $y$, omitting oscillating terms, we
then find
\begin{eqnarray}
  I_2 &=& \frac{\lambda \tau_{\rm D}}{\pi^2} \int_0^{1/2} dx' (x'^{\epsilon} -
  (1/2)^{\epsilon}) \frac{\cos(r x')}{1-x'} \nonumber \\ && \mbox{}
  - \frac{F_6(1,2,3,4,5,6)}{\pi^2}
  \int_0^{1/2} dx' (x'^{\epsilon} - (1/2)^{\epsilon})
  \int_0^1 \frac{dy}{y} y^{\eta}
  \frac{x' \cos(r y x') - \cos(r y) + (1 - x') \cos(y r (1-x')]}{x'(1-x')}
  \nonumber \\ && \mbox{}
  + \ldots.
\end{eqnarray}
The integral in the first line is of order $1/r$ and can be neglected
in the limit $r \to \infty$. The integral in the second line is of
order $\epsilon \propto 1/\lambda \tau_{\rm D}$, so that it can also be
neglected. This is best seen by evaluating the integral after
expanding $(x')^{\epsilon} - (1/2)^{\epsilon} \approx \epsilon \ln(2 x')$
and replacing $y^{\eta}$ by $1$, which gives a finite value of the
integral. Hence, we conclude that, in the classical limit $r \to
\infty$ at fixed $r^{-1/\lambda \tau_{\rm D}}$, we have
\begin{equation}
  I_2 = 0.
  \label{eq:I22}
\end{equation}
Similarly, one finds that the third term in Eq.\ (\ref{eq:su}) does
not contribute to $I$.

For the remaining contribution $I_4$, we note that the only nonzero
contribution can come from small $x'$ and small $y'$. Neglecting $x'$ and
$y'$ with respect to unity, one finds that the contribution from the
first term between brackets $\{ \ldots \}$ in Eq.\ (\ref{eq:F2six})
gives zero. In the same approximation, after two partial integrations
the second term gives
\begin{eqnarray}
  I_4 &=& \frac{4 \lambda \tau_{\rm D} \epsilon_1 \epsilon_2}{\pi^2}
  \int_0^{1/2} \frac{dx' dy'}{x' y'} \int_0^1 dy y^{\eta-1}
  (x')^{\epsilon_1}
  (y')^{\epsilon_2}
  \sin(y r x') \sin(y r y') + \ldots,
\end{eqnarray}
where $\epsilon_1 = F_4(1,2,3,4)/\lambda \tau_{\rm D}$ and $\epsilon_2
= F_4(5,4,1,6)$ (or cyclic permutations), and $\eta =
F_6(1,2,3,4,5,6)/\lambda \tau_{\rm D}$. We then shift variables $x' =
x''/y$ and $y' = y''/y$ and integrate over $y$,
\begin{eqnarray} \label{eq:I4}
  I_4 &=& \frac{4 \lambda \tau_{\rm D} \epsilon_1 \epsilon_2}
  {\pi^2(\epsilon_1+\epsilon_2-\eta)}
  \int_0^{1/2} \frac{dx'' dy''}{x'' y''}
  (x'')^{\epsilon_1} (y'')^{\epsilon_1}
  [(\max(x'',y''))^{\eta-\epsilon_1-\epsilon_2} -
  (1/2)^{\eta-\epsilon_1-\epsilon_2}]
  \sin(r x'') \sin(r y'') + \ldots \nonumber \\
  &=&
  \frac{F_4(1,2,3,4) F_2(5,4,1,6)}{F_2(1,4)}
  e^{-\tau_{\rm E} F_6(1,2,3,4,5,6)/\tau_{\rm D}}
  \left[1
  - e^{- \tau_{\rm E} F_2(1,4)/\tau_{\rm D}}
  \right]
  + \ldots,
\end{eqnarray}
\end{widetext}
where we used that $F_4(1,2,3,4) + F_4(5,4,1,6) - F_6(1,2,3,4,5,6) =
F_2(1,4)$. As before, the dots \ldots refer to the two additional
terms obtained by cyclic permutation $(1,2) \to (3,4) \to (5,6)$
of the expression written above.

The final result for $Q_6^{(\ref{fig:3}b)}$ is found by 
substituting $I = I_1 + I_4$ into Eq.\ (\ref{eq:A3b}).

\subsection{Three-encounter that touches the lead openings}
\label{sec:sub2}

If the three-encounter touches the lead openings, escape of the
trajectories involved in the encounter is no longer uncorrelated. In
order to deal with three-encounters that touch the lead openings, the
initial part of the calculation of the previous subsection has to be
modified. However, the final integrals will all be of the form of the
integrals $I_1$, $I_2$, and $I_4$ calculated above. Here we discuss
the configurations of Figs.\ \ref{fig:4}b and c in detail. The 
calculation of the remaining 
configurations (Figs.\ \ref{fig:8}e--g) proceeds in 
a similar manner. In the discussion here
we further limit ourselves to the labeling of trajectories as shown in 
the figure. For Fig.\ \ref{fig:4}b, this means that we consider the 
exit of
trajectories $\alpha_1$ and $\alpha_5$ to be correlated, whereas the exit
of trajectory $\alpha_3$, as well as all entrances are
uncorrelated. There are five more contributions to $Q_6$: two of these
arise
from configurations where the exit of $\alpha_1$ and $\alpha_3$ or the
exit of $\alpha_3$ and $\alpha_5$ is correlated; three more arise from
configurations with correlated entry of two trajectories, and
uncorrelated exits. For Fig.\ \ref{fig:4}c, we take the exits of all
trajectories to be correlated, whereas the entrances are not. There is
another contribution obtained by reversing the roles of entrance and
exit.


In order to describe a three-encounter that touches the lead opening
as shown in Fig.\ \ref{fig:4}b, 
we again take a Poincar\'e surface of section at a point that all
six trajectories are within a phase space distance $c$. In the
configuration of Fig.\ \ref{fig:4}b, this `central' part of the
three-encounter does not touch a lead opening. 
\begin{figure}
\epsfxsize=0.6\hsize
\hspace{0.05\hsize}
\epsffile{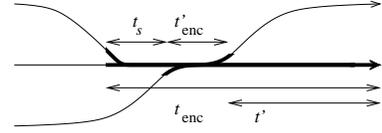}
\caption{\label{fig:enctimes2}
  Definition of the four measures of the encounter duration
  used in the text: $t_{\rm enc}$ is the total duration of the
  three-encounter, $t_{\rm enc}'$ is the duration of the part of the
  encounter that all trajectories involved in the three-encounter are
  within a phase-space distance $c$, and $t_s$ and $t'$ are the
  durations of the encounter stretches where one pair of trajectories
  has already diverged from the remaining two pairs. The right end of
  the encounter is at the lead opening.}
\end{figure}
We ensure that $\alpha_3$ has moved away from the other
trajectories before arriving at the lead opening by requiring 
\begin{equation}
  \min(|u_1-u_3|,|u_5-u_3|) > |u_1-u_5|.
\end{equation}
The escape of
$\alpha_1$ and $\alpha_5$ is correlated if
\begin{equation}
  0 < t' < t_u,
\end{equation}
where
\begin{equation}
  t_u = \frac{1}{\lambda} \ln \frac{\max(|u_3-u_1|,|u_3-u_5|)}{|u_5-u_1|},
\end{equation}
and $t'$
is the duration of the stretch of the encounter immediately adjacent
to the lead opening, during which $\alpha_1$ and $\alpha_5$ are
correlated with each other, but not with $\alpha_3$, see Fig.\
\ref{fig:enctimes2}.
We then find
\begin{eqnarray}
  \label{eq:P6bI}
  Q_6^{(\ref{fig:4}b)} &=&
  \frac{N_i N_j N_k N_l N_n \delta_{im}}{N^5 F_2(1,2) F_2(3,4)
  F_2(5,6) F_2(3,2)}
  I, ~~~
\end{eqnarray}
where
\begin{eqnarray}
  I &=& \int d(s_3-s_1) d(s_5-s_1) d(u_3-u_1) d(u_5-u_1)
  \nonumber \\ && \mbox{} \times
  \int \frac{dt'}{\tau_{\rm D}}
  \frac{  e^{i \Delta S_{\rm enc} - t_{\rm enc}/\tau_{\rm D}}}{(2 \pi \hbar)^2 t_{\rm enc}'},
\end{eqnarray}
with
\begin{eqnarray}
  t_{\rm enc} &=& t' + t_{\rm enc}' + t_s, \\
  t_{\rm enc}' &=& \frac{1}{\lambda} \ln \frac{c}{\max(|u_3-u_1|,|u_3-u_5|)}
  \nonumber \\ && \mbox{}
  +  \frac{1}{\lambda} \ln
  \frac{c}{\max(|s_1-s_3|,|s_1-s_5|,|s_3-s_5|)}. ~~~ \nonumber \\
  t_s &=& \frac{1}{\lambda} \ln
  \frac{\max(|s_1-s_3|,|s_1-s_5|,|s_3-s_5|)}
  {\min(|s_1-s_3|,|s_1-s_5|,|s_3-s_5|)}. 
\end{eqnarray}
The action difference $\Delta {\cal S}_{\rm enc}$ is given by Eq.\
(\ref{eq:Sencdef}) above.
  
Rewriting the integral in terms of the variables $s =
\max(|s_1-s_3|,|s_1-s_5|,|s_3-s_5|)$, $s' =
\min(|s_1-s_3|,|s_1-s_5|,|s_3-s_5|)$, $u = \max(|u_3-u_1|,|u_3-u_5|)$,
and $u' = |u_5-u_1|$, and integrating over $t'$, we arrive at
\begin{widetext}
\begin{eqnarray}
  I &=& \tau_{\rm D}
  \int_0^{c} ds \int_0^{s/2} ds' \int_0^{c} du \int_{0}^{u/2} du'
  \frac{ e^{-t_{\rm enc}' F_6(1,2,3,4,5,6)/\tau_{\rm D}}}
  {(\pi \hbar)^2 t_{\rm enc}' F_4(5,4,1,6)}
  \left[1 - e^{-t_u F_4(1,4,5,6)/\tau_{\rm D}}
  \right]
  \nonumber \\ && \mbox{} \times
  \left\{ (
  e^{-t_s F_4(1,2,5,6)/\tau_{\rm D}} 
  + e^{-t_s F_4(3,4,5,6)/\tau_{\rm D}} )
  \right. \nonumber \\ && \left. \ \ \ \ \mbox{} \times
  \left( \cos \frac{u s + u' s' - u' s}{\hbar} +
  \cos \frac{u s - u s' - u' s}{\hbar} +
  \cos \frac{us - u's'}{\hbar} +
  \cos \frac{us - us' + u's'}{\hbar} \right) 
  \right. \nonumber \\ && \ \ \ \ \left. \mbox{} +
  2 e^{-t_s F_4(1,2,3,4)/\tau_{\rm D}}
  \left(\cos \frac{u s' - u' s}{\hbar} +
  \cos \frac{u s' + u' s - u s'}{\hbar} \right) \right\},
\end{eqnarray}
We now change variables
\begin{eqnarray}
  s = c/\sigma,\ \ s' = c x'/\sigma, \ \
  u = c y \sigma, \ \ u' = c y y' \sigma.
\end{eqnarray}
Writing $r = c^2/\hbar$ and integrating over $\sigma$, the integral then reads
\begin{eqnarray}
  \label{eq:If}
  I &=& \frac{\lambda \tau_{\rm D} r^2}{\pi^2 F_4(5,4,1,6)}
  \int_0^{1/2} dx' \int_{0}^{1/2}
  dy' \int_0^1 y dy y^{F_6(1,2,3,4,5,6)/\lambda
  \tau_{\rm D}}
  [1 - (y')^{F_4(1,4,5,6)/\lambda \tau_{\rm D}}]
  \nonumber \\ && \mbox{} \times
  \left\{
  [(x')^{F_4(1,2,5,6)/\lambda \tau_{\rm D}} 
  + (x')^{F_4(3,4,5,6)/\lambda \tau_{\rm D}}]
  \right. \nonumber \\ && \ \ \ \ \left. \mbox{} \times
  [\cos(y r(1 + y' x' - y')) + \cos(y r(1 - x' - y')
  + \cos(yr(1 - x' y')) + \cos(yr(1-x' + y' x'))]
  \right. \nonumber \\ && \ \ \ \ \left. \mbox{}
  + 2 (x')^{F_4(1,2,3,4)/\lambda \tau_{\rm D}}
  [\cos(y r(x'+y' - x'y')) + \cos(y r(x'-y'))] \right\}.
\end{eqnarray}
The remainder of the calculation is identical to that of the previous
subsection: We write 
\begin{equation}
  (x')^{\epsilon} = 1 + ((x')^{\epsilon} - (1/2)^{\epsilon}),
  \label{eq:su2}
\end{equation}
where $\epsilon$ represents $F_4(1,2,5,6)/\lambda \tau_{\rm D}$,
$F_4(3,4,5,6)/\lambda \tau_{\rm D}$, or $F_4(1,2,3,4)/\lambda
\tau_{\rm D}$, and evaluate the two terms in Eq.\ (\ref{eq:su2})
separately. The calculation of the first term is identical to that of
$I_2$, see Eqs.\ (\ref{eq:I21})--(\ref{eq:I22}), and 
vanishes. The calculation of the second term is identical to that of
$I_4$, see Eq.\ (\ref{eq:I4}) above. We thus find
\begin{eqnarray}
  \label{eq:I6b}
  I &=& -\frac{F_4(1,2,3,4)}{F_2(1,4)}
  e^{-\tau_{\rm E} F_6(1,2,3,4,5,6)/\tau_{\rm D}}
  \left[ 1 - e^{-\tau_{\rm E} F_2(1,4)/\tau_{\rm D}} \right].
\end{eqnarray}
\end{widetext}
Substitution of Eq. (\ref{eq:I6b}) into Eq.\ (\ref{eq:P6bI}) 
and addition of the remaining five configurations of trajectories obtained
by relabeling or by interchanging entrance and exit gives 
Eq.\ (\ref{eq:P6b}).

Proceeding in a similar manner for the case shown in Fig.\
\ref{fig:7}c, we find
\begin{eqnarray}
  Q_6^{\ref{fig:7}c} 
  &=& \frac{N_i N_j N_l N_n \delta_{ik} \delta_{im}}{N^4
  F_2(1,2) F_2(3,4) F_2(5,6)} I,
  \label{eq:P67I}
\end{eqnarray}
where 
\begin{eqnarray}
  I &=& \tau_{\rm D} \int d(s_3-s_1) d(s_5-s_1) d(u_3-u_1)
    d(u_5-u_1)
  \nonumber \\ && \mbox{} \times
  \int \frac{dt_{\rm enc}'}{\tau_{\rm D}}
  \frac{e^{i \Delta S_{\rm enc}/\hbar - t_{\rm enc}/\tau_{\rm D}}}
  {(2 \pi \hbar)^2 t_{\rm enc}'},
\end{eqnarray}
with
\begin{eqnarray}
  t_{\rm enc} &=& t_{\rm enc}' + \frac{1}{\lambda}
  \ln
  \frac{\max(|s_1-s_3|,|s_1-s_5|,|s_3-s_5|)}
    {\min(|s_1-s_3|,|s_1-s_5|,|s_3-s_5|)}.
  \nonumber \\
\end{eqnarray}
We require
\begin{eqnarray}
  0 < t_{\rm enc}' &<& \frac{1}{\lambda} \ln
  \frac{c}{\max(|s_1-s_3|,|s_1-s_5|,|s_3-s_5|)}
  \nonumber \\ && \mbox{}
  + \frac{1}{\lambda} \ln
  \frac{c}{\max(|u_1-u_3|,|u_1-u_5|,|u_3-u_5|)},
  \nonumber \\
\end{eqnarray}
in order to ensure that the three-encounter indeed touches the lead
opening at a point that all trajectories involved in the encounter are
separated by phase space distances less than the cut-off $c$.
The integration over $t_{\rm enc}'$ can be done immediately. Rewriting the
integral in terms of the variables $s =
\max(|s_1-s_3|,|s_1-s_5|,|s_3-s_5|)$, $s' =
\min(|s_1-s_3|,|s_1-s_5|,|s_3-s_5|)$, $u =
\max(|u_1-u_3|,|u_1-u_5|,|u_3-u_5|)$, and $u' =
\min(|u_1-u_3|,|u_1-u_5|,|u_3-u_5|)$, and performing the variable change
\begin{eqnarray}
  s = c/\sigma,\ \ s' = c x'/\sigma, \ \
  u = c y \sigma, \ \ u' = c y y' \sigma,
\end{eqnarray}
we arrive at the integral
\begin{widetext}
\begin{eqnarray}
  I &=&\frac{2 \lambda \tau_{\rm D} r^2}{\pi^2 F_6(1,2,3,4,5,6)}
  \int_0^{1/2} dx' dy' \int_0^1 y dy
  [1 - y^{F_6(1,2,3,4,5,6)/\lambda \tau_{\rm D}}]
  \left[ (x')^{F_4(1,2,5,6)/\lambda \tau_{\rm D}} 
  + \ldots \right]
  \nonumber \\ && \mbox{} \times
  [\cos(y r(1 + y' x' - y')) + \cos(y r(1 - x' - y')
  +
  \cos(yr(1 - x' y')) + \cos(yr(1-x' + y' x'))
  \nonumber \\ && \ \ \ \ \mbox{} 
  + 
  \cos(y r(x'+y' - x'y')) + \cos(y r(x'-y'))].
\end{eqnarray}
\end{widetext}
In this integral, the contribution from the term $1$ in $[1 -
  y^{F_6(1,2,3,4,5,6)/\lambda \tau_{\rm D}}]$ vanishes. Writing
$
  (x')^{\epsilon} = (1/2)^{\epsilon} + ((x')^{\epsilon} - 
(1/2)^{\epsilon}),
$
with $\epsilon = F_4(1,2,5,6)/\lambda \tau_{\rm D}$ etc., we arrive
at integrals identical to the integrals $I_1$ and $I_2$ considered in
the previous subsection. We thus find
\begin{eqnarray}
  I &=& e^{-\tau_{\rm E} F_6(1,2,3,4,5,6)/\tau_{\rm D}}.
\end{eqnarray}
Substituting this result into Eq.\ (\ref{eq:P67I}) and adding what one
obtains after interchanging entrance and exit leads, one arrives at
Eq.\ (\ref{eq:P67}).

The remaining configurations of trajectories with a three-encounter
(Figs.\ \ref{fig:8}e and \ref{fig:10}) are treated in the same manner.


\begin{thebibliography}{42}
\expandafter\ifx\csname natexlab\endcsname\relax\def\natexlab#1{#1}\fi
\expandafter\ifx\csname bibnamefont\endcsname\relax
  \def\bibnamefont#1{#1}\fi
\expandafter\ifx\csname bibfnamefont\endcsname\relax
  \def\bibfnamefont#1{#1}\fi
\expandafter\ifx\csname citenamefont\endcsname\relax
  \def\citenamefont#1{#1}\fi
\expandafter\ifx\csname url\endcsname\relax
  \def\url#1{\texttt{#1}}\fi
\expandafter\ifx\csname urlprefix\endcsname\relax\def\urlprefix{URL }\fi
\providecommand{\bibinfo}[2]{#2}
\providecommand{\eprint}[2][]{\url{#2}}

\bibitem[{\citenamefont{Aleiner and Larkin}(1996)}]{kn:aleiner1996}
\bibinfo{author}{\bibfnamefont{I.~L.} \bibnamefont{Aleiner}} \bibnamefont{and}
  \bibinfo{author}{\bibfnamefont{A.~I.} \bibnamefont{Larkin}},
  \bibinfo{journal}{Phys. Rev. B} \textbf{\bibinfo{volume}{54}},
  \bibinfo{pages}{14423} (\bibinfo{year}{1996}).

\bibitem[{\citenamefont{Larkin and Ovchinnikov}(1968)}]{kn:larkin1968}
\bibinfo{author}{\bibfnamefont{A.~I.} \bibnamefont{Larkin}} \bibnamefont{and}
  \bibinfo{author}{\bibfnamefont{Y.~N.} \bibnamefont{Ovchinnikov}},
  \bibinfo{journal}{Zh. Eksp. Teor. Fiz.} \textbf{\bibinfo{volume}{55}},
  \bibinfo{pages}{2262} (\bibinfo{year}{1968}) \bibinfo{note}{[Sov.\ Phys.\
  JETP {\bf 28}, 1200 (1969)]}.

\bibitem[{\citenamefont{Zaslavsky}(1981)}]{kn:zaslavsky1981}
\bibinfo{author}{\bibfnamefont{G.~M.} \bibnamefont{Zaslavsky}},
  \bibinfo{journal}{Phys. Rep.} \textbf{\bibinfo{volume}{80}},
  \bibinfo{pages}{157} (\bibinfo{year}{1981}).

\bibitem[{\citenamefont{Kouwenhoven et~al.}(1997)\citenamefont{Kouwenhoven,
  Marcus, McEuen, Tarucha, Westervelt, and Wingreen}}]{kn:kouwenhoven1997}
\bibinfo{author}{\bibfnamefont{L.~P.} \bibnamefont{Kouwenhoven}},
  \bibinfo{author}{\bibfnamefont{C.~M.} \bibnamefont{Marcus}},
  \bibinfo{author}{\bibfnamefont{P.~L.} \bibnamefont{McEuen}},
  \bibinfo{author}{\bibfnamefont{S.}~\bibnamefont{Tarucha}},
  \bibinfo{author}{\bibfnamefont{R.~M.} \bibnamefont{Westervelt}},
  \bibnamefont{and} \bibinfo{author}{\bibfnamefont{N.~S.}
  \bibnamefont{Wingreen}}, in \emph{\bibinfo{booktitle}{Mesoscopic Electron
  Transport}}, edited by \bibinfo{editor}{\bibfnamefont{L.~L.}
  \bibnamefont{Sohn}}, \bibinfo{editor}{\bibfnamefont{L.~P.}
  \bibnamefont{Kouwenhoven}}, \bibnamefont{and}
  \bibinfo{editor}{\bibfnamefont{G.}~\bibnamefont{Sch\"on}}
  (\bibinfo{publisher}{Kluwer, Dordrecht}, \bibinfo{year}{1997}), vol.
  \bibinfo{volume}{345} of \emph{\bibinfo{series}{NATO ASI Series E}}.

\bibitem[{\citenamefont{Beenakker}(1997)}]{kn:beenakker1997}
\bibinfo{author}{\bibfnamefont{C.~W.~J.} \bibnamefont{Beenakker}},
  \bibinfo{journal}{Rev. Mod. Phys.} \textbf{\bibinfo{volume}{69}},
  \bibinfo{pages}{731} (\bibinfo{year}{1997}).

\bibitem[{\citenamefont{Agam et~al.}(2000)\citenamefont{Agam, Aleiner, and
  Larkin}}]{kn:agam2000}
\bibinfo{author}{\bibfnamefont{O.}~\bibnamefont{Agam}},
  \bibinfo{author}{\bibfnamefont{I.}~\bibnamefont{Aleiner}}, \bibnamefont{and}
  \bibinfo{author}{\bibfnamefont{A.}~\bibnamefont{Larkin}},
  \bibinfo{journal}{Phys. Rev. Lett.} \textbf{\bibinfo{volume}{85}},
  \bibinfo{pages}{3153} (\bibinfo{year}{2000}).

\bibitem[{\citenamefont{Silvestrov
  et~al.}(2003{\natexlab{a}})\citenamefont{Silvestrov, Goorden, and
  Beenakker}}]{kn:silvestrov2003}
\bibinfo{author}{\bibfnamefont{P.~G.} \bibnamefont{Silvestrov}},
  \bibinfo{author}{\bibfnamefont{M.~C.} \bibnamefont{Goorden}},
  \bibnamefont{and} \bibinfo{author}{\bibfnamefont{C.~W.~J.}
  \bibnamefont{Beenakker}}, \bibinfo{journal}{Phys. Rev. B}
  \textbf{\bibinfo{volume}{67}}, \bibinfo{eid}{241301}
  (\bibinfo{year}{2003}{\natexlab{a}}).

\bibitem[{\citenamefont{Tworzydlo et~al.}(2003)\citenamefont{Tworzydlo, Tajic,
  Schomerus, and Beenakker}}]{kn:tworzydlo2003}
\bibinfo{author}{\bibfnamefont{J.}~\bibnamefont{Tworzydlo}},
  \bibinfo{author}{\bibfnamefont{A.}~\bibnamefont{Tajic}},
  \bibinfo{author}{\bibfnamefont{H.}~\bibnamefont{Schomerus}},
  \bibnamefont{and} \bibinfo{author}{\bibfnamefont{C.~W.~J.}
  \bibnamefont{Beenakker}}, \bibinfo{journal}{Phys. Rev. B}
  \textbf{\bibinfo{volume}{68}}, \bibinfo{eid}{115313} (\bibinfo{year}{2003}).

\bibitem[{\citenamefont{Whitney and Jacquod}(2005)}]{kn:whitney2005}
\bibinfo{author}{\bibfnamefont{R.~S.} \bibnamefont{Whitney}} \bibnamefont{and}
  \bibinfo{author}{\bibfnamefont{P.}~\bibnamefont{Jacquod}},
  \bibinfo{journal}{Phys. Rev. Lett.} \textbf{\bibinfo{volume}{94}},
  \bibinfo{eid}{116801} (\bibinfo{year}{2005}).

\bibitem[{\citenamefont{Adagideli and Beenakker}(2002)}]{kn:adagideli2002}
\bibinfo{author}{\bibfnamefont{I.}~\bibnamefont{Adagideli}} \bibnamefont{and}
  \bibinfo{author}{\bibfnamefont{C.~W.~J.} \bibnamefont{Beenakker}},
  \bibinfo{journal}{Phys. Rev. Lett.} \textbf{\bibinfo{volume}{89}},
  \bibinfo{pages}{237002} (\bibinfo{year}{2002}).

\bibitem[{\citenamefont{Vavilov and Larkin}(2003)}]{kn:vavilov2003}
\bibinfo{author}{\bibfnamefont{M.~G.} \bibnamefont{Vavilov}} \bibnamefont{and}
  \bibinfo{author}{\bibfnamefont{A.~I.} \bibnamefont{Larkin}},
  \bibinfo{journal}{Phys. Rev. B} \textbf{\bibinfo{volume}{67}},
  \bibinfo{eid}{115335} (\bibinfo{year}{2003}).

\bibitem[{\citenamefont{Silvestrov
  et~al.}(2003{\natexlab{b}})\citenamefont{Silvestrov, Goorden, and
  Beenakker}}]{kn:silvestrov2003b}
\bibinfo{author}{\bibfnamefont{P.~G.} \bibnamefont{Silvestrov}},
  \bibinfo{author}{\bibfnamefont{M.~C.} \bibnamefont{Goorden}},
  \bibnamefont{and} \bibinfo{author}{\bibfnamefont{C.~W.~J.}
  \bibnamefont{Beenakker}}, \bibinfo{journal}{Phys. Rev. Lett.}
  \textbf{\bibinfo{volume}{90}}, \bibinfo{eid}{116801}
  (\bibinfo{year}{2003}{\natexlab{b}}).

\bibitem[{\citenamefont{Beenakker}(2005)}]{kn:beenakker2004}
\bibinfo{author}{\bibfnamefont{C.~W.~J.} \bibnamefont{Beenakker}},
  \bibinfo{journal}{Lect. Notes Phys.} \textbf{\bibinfo{volume}{667}},
  \bibinfo{pages}{131} (\bibinfo{year}{2005}),
  \bibinfo{note}{cond-mat/0406018}.

\bibitem[{\citenamefont{Adagideli}(2003)}]{kn:adagideli2003}
\bibinfo{author}{\bibfnamefont{I.}~\bibnamefont{Adagideli}},
  \bibinfo{journal}{Phys. Rev. B} \textbf{\bibinfo{volume}{68}},
  \bibinfo{eid}{233308} (\bibinfo{year}{2003}).

\bibitem[{\citenamefont{Rahav and Brouwer}(2005)}]{kn:rahav2005}
\bibinfo{author}{\bibfnamefont{S.}~\bibnamefont{Rahav}} \bibnamefont{and}
  \bibinfo{author}{\bibfnamefont{P.~W.} \bibnamefont{Brouwer}},
  \bibinfo{journal}{Phys.\ Rev.\ Lett.} \textbf{\bibinfo{volume}{95}},
  \bibinfo{eid}{056806} (\bibinfo{year}{2005}).

\bibitem[{\citenamefont{Jacquod and Whitney}(2005)}]{kn:jacquod2006}
\bibinfo{author}{\bibfnamefont{P.}~\bibnamefont{Jacquod}} \bibnamefont{and}
  \bibinfo{author}{\bibfnamefont{R.~S.} \bibnamefont{Whitney}},
  \bibinfo{journal}{cond-mat/0512662}  (\bibinfo{year}{2005}).

\bibitem[{\citenamefont{Tworzydlo
  et~al.}(2004{\natexlab{a}})\citenamefont{Tworzydlo, Tajic, and
  Beenakker}}]{kn:tworzydlo2004}
\bibinfo{author}{\bibfnamefont{J.}~\bibnamefont{Tworzydlo}},
  \bibinfo{author}{\bibfnamefont{A.}~\bibnamefont{Tajic}}, \bibnamefont{and}
  \bibinfo{author}{\bibfnamefont{C.~W.~J.} \bibnamefont{Beenakker}},
  \bibinfo{journal}{Phys. Rev. B} \textbf{\bibinfo{volume}{69}},
  \bibinfo{eid}{165318} (\bibinfo{year}{2004}{\natexlab{a}}).

\bibitem[{\citenamefont{Jacquod and Sukhorukov}(2004)}]{kn:jacquod2004}
\bibinfo{author}{\bibfnamefont{P.}~\bibnamefont{Jacquod}} \bibnamefont{and}
  \bibinfo{author}{\bibfnamefont{E.~V.} \bibnamefont{Sukhorukov}},
  \bibinfo{journal}{Phys. Rev. Lett.} \textbf{\bibinfo{volume}{92}},
  \bibinfo{eid}{116801} (\bibinfo{year}{2004}).

\bibitem[{\citenamefont{Brouwer and Rahav}(2005)}]{kn:brouwer2006}
\bibinfo{author}{\bibfnamefont{P.~W.} \bibnamefont{Brouwer}} \bibnamefont{and}
  \bibinfo{author}{\bibfnamefont{S.}~\bibnamefont{Rahav}},
  \bibinfo{journal}{cond-mat/0512095}  (\bibinfo{year}{2005}).

\bibitem[{\citenamefont{Aleiner and Larkin}(1997)}]{kn:aleiner1997}
\bibinfo{author}{\bibfnamefont{I.~L.} \bibnamefont{Aleiner}} \bibnamefont{and}
  \bibinfo{author}{\bibfnamefont{A.~I.} \bibnamefont{Larkin}},
  \bibinfo{journal}{Phys. Rev. E} \textbf{\bibinfo{volume}{55}},
  \bibinfo{pages}{1243(R)} (\bibinfo{year}{1997}).

\bibitem[{\citenamefont{Tian and Larkin}(2004)}]{kn:tian2004b}
\bibinfo{author}{\bibfnamefont{C.}~\bibnamefont{Tian}} \bibnamefont{and}
  \bibinfo{author}{\bibfnamefont{A.~I.} \bibnamefont{Larkin}},
  \bibinfo{journal}{Phys. Rev. B} \textbf{\bibinfo{volume}{70}},
  \bibinfo{eid}{035305} (\bibinfo{year}{2004}).

\bibitem[{\citenamefont{Tworzydlo
  et~al.}(2004{\natexlab{b}})\citenamefont{Tworzydlo, Tajic, Schomerus,
  Brouwer, and Beenakker}}]{kn:tworzydlo2004c}
\bibinfo{author}{\bibfnamefont{J.}~\bibnamefont{Tworzydlo}},
  \bibinfo{author}{\bibfnamefont{A.}~\bibnamefont{Tajic}},
  \bibinfo{author}{\bibfnamefont{H.}~\bibnamefont{Schomerus}},
  \bibinfo{author}{\bibfnamefont{P.~W.} \bibnamefont{Brouwer}},
  \bibnamefont{and} \bibinfo{author}{\bibfnamefont{C.~W.~J.}
  \bibnamefont{Beenakker}}, \bibinfo{journal}{Phys. Rev. Lett.}
  \textbf{\bibinfo{volume}{93}}, \bibinfo{eid}{186806}
  (\bibinfo{year}{2004}{\natexlab{b}}).

\bibitem[{\citenamefont{Goorden et~al.}(2005)\citenamefont{Goorden, Jacquod,
  and Beenakker}}]{kn:goorden2005}
\bibinfo{author}{\bibfnamefont{M.~C.} \bibnamefont{Goorden}},
  \bibinfo{author}{\bibfnamefont{P.}~\bibnamefont{Jacquod}}, \bibnamefont{and}
  \bibinfo{author}{\bibfnamefont{C.~W.~J.} \bibnamefont{Beenakker}},
  \bibinfo{journal}{Phys.\ Rev.\ B} \textbf{\bibinfo{volume}{72}},
  \bibinfo{eid}{064526} (\bibinfo{year}{2005}).

\bibitem[{\citenamefont{Schomerus and Jacquod}(2005)}]{kn:schomerus2005}
\bibinfo{author}{\bibfnamefont{H.}~\bibnamefont{Schomerus}} \bibnamefont{and}
  \bibinfo{author}{\bibfnamefont{P.}~\bibnamefont{Jacquod}},
  \bibinfo{journal}{J. Phys. A} \textbf{\bibinfo{volume}{38}},
  \bibinfo{pages}{10663} (\bibinfo{year}{2005}).

\bibitem[{\citenamefont{Schomerus and Tworzydlo}(2004)}]{kn:schomerus2004}
\bibinfo{author}{\bibfnamefont{H.}~\bibnamefont{Schomerus}} \bibnamefont{and}
  \bibinfo{author}{\bibfnamefont{J.}~\bibnamefont{Tworzydlo}},
  \bibinfo{journal}{Phys. Rev. Lett.} \textbf{\bibinfo{volume}{93}},
  \bibinfo{eid}{154102} (\bibinfo{year}{2004}).

\bibitem[{\citenamefont{Blanter and B\"uttiker}(2000)}]{kn:blanter2000b}
\bibinfo{author}{\bibfnamefont{Y.~M.} \bibnamefont{Blanter}} \bibnamefont{and}
  \bibinfo{author}{\bibfnamefont{M.}~\bibnamefont{B\"uttiker}},
  \bibinfo{journal}{Phys. Rep.} \textbf{\bibinfo{volume}{336}},
  \bibinfo{pages}{1} (\bibinfo{year}{2000}).

\bibitem[{\citenamefont{Bl\"umel and Smilansky}(1988)}]{kn:bluemel1988}
\bibinfo{author}{\bibfnamefont{R.}~\bibnamefont{Bl\"umel}} \bibnamefont{and}
  \bibinfo{author}{\bibfnamefont{U.}~\bibnamefont{Smilansky}},
  \bibinfo{journal}{Phys. Rev. Lett.} \textbf{\bibinfo{volume}{60}},
  \bibinfo{pages}{477} (\bibinfo{year}{1988}).

\bibitem[{\citenamefont{Polianski and Brouwer}(2003)}]{kn:polianski2003}
\bibinfo{author}{\bibfnamefont{M.~L.} \bibnamefont{Polianski}}
  \bibnamefont{and} \bibinfo{author}{\bibfnamefont{P.~W.}
  \bibnamefont{Brouwer}}, \bibinfo{journal}{J. Phys. A}
  \textbf{\bibinfo{volume}{36}}, \bibinfo{pages}{3215} (\bibinfo{year}{2003}).

\bibitem[{\citenamefont{Brouwer et~al.}(2005)\citenamefont{Brouwer, Lamacraft,
  and Flensberg}}]{kn:brouwer2005d}
\bibinfo{author}{\bibfnamefont{P.~W.} \bibnamefont{Brouwer}},
  \bibinfo{author}{\bibfnamefont{A.}~\bibnamefont{Lamacraft}},
  \bibnamefont{and}
  \bibinfo{author}{\bibfnamefont{K.}~\bibnamefont{Flensberg}},
  \bibinfo{journal}{Phys. Rev. B} \textbf{\bibinfo{volume}{72}},
  \bibinfo{eid}{075316} (\bibinfo{year}{2005}).

\bibitem[{\citenamefont{Jalabert et~al.}(1990)\citenamefont{Jalabert, Baranger,
  and Stone}}]{kn:jalabert1990}
\bibinfo{author}{\bibfnamefont{R.~A.} \bibnamefont{Jalabert}},
  \bibinfo{author}{\bibfnamefont{H.~U.} \bibnamefont{Baranger}},
  \bibnamefont{and} \bibinfo{author}{\bibfnamefont{A.~D.} \bibnamefont{Stone}},
  \bibinfo{journal}{Phys. Rev. Lett.} \textbf{\bibinfo{volume}{65}},
  \bibinfo{pages}{2442} (\bibinfo{year}{1990}).

\bibitem[{\citenamefont{Baranger
  et~al.}(1993{\natexlab{a}})\citenamefont{Baranger, Jalabert, and
  Stone}}]{kn:baranger1993}
\bibinfo{author}{\bibfnamefont{H.~U.} \bibnamefont{Baranger}},
  \bibinfo{author}{\bibfnamefont{R.~A.} \bibnamefont{Jalabert}},
  \bibnamefont{and} \bibinfo{author}{\bibfnamefont{A.~D.} \bibnamefont{Stone}},
  \bibinfo{journal}{Phys. Rev. Lett.} \textbf{\bibinfo{volume}{70}},
  \bibinfo{pages}{3876} (\bibinfo{year}{1993}{\natexlab{a}}).

\bibitem[{\citenamefont{Baranger
  et~al.}(1993{\natexlab{b}})\citenamefont{Baranger, Jalabert, and
  Stone}}]{kn:baranger1993b}
\bibinfo{author}{\bibfnamefont{H.~U.} \bibnamefont{Baranger}},
  \bibinfo{author}{\bibfnamefont{R.~A.} \bibnamefont{Jalabert}},
  \bibnamefont{and} \bibinfo{author}{\bibfnamefont{A.~D.} \bibnamefont{Stone}},
  \bibinfo{journal}{Chaos} \textbf{\bibinfo{volume}{3}}, \bibinfo{pages}{665}
  (\bibinfo{year}{1993}{\natexlab{b}}).

\bibitem[{\citenamefont{Braun et~al.}(2005)\citenamefont{Braun, Heusler,
  M\"uller, and Haake}}]{kn:braun2005}
\bibinfo{author}{\bibfnamefont{P.}~\bibnamefont{Braun}},
  \bibinfo{author}{\bibfnamefont{S.}~\bibnamefont{Heusler}},
  \bibinfo{author}{\bibfnamefont{S.}~\bibnamefont{M\"uller}}, \bibnamefont{and}
  \bibinfo{author}{\bibfnamefont{F.}~\bibnamefont{Haake}},
  \bibinfo{journal}{cond-mat/051192}  (\bibinfo{year}{2005}).

\bibitem[{\citenamefont{Heusler et~al.}(2006)\citenamefont{Heusler, M\"uller,
  Braun, and Haake}}]{kn:heusler2006}
\bibinfo{author}{\bibfnamefont{S.}~\bibnamefont{Heusler}},
  \bibinfo{author}{\bibfnamefont{S.}~\bibnamefont{M\"uller}},
  \bibinfo{author}{\bibfnamefont{P.}~\bibnamefont{Braun}}, \bibnamefont{and}
  \bibinfo{author}{\bibfnamefont{F.}~\bibnamefont{Haake}},
  \bibinfo{journal}{Phys. Rev. Lett.} \textbf{\bibinfo{volume}{96}},
  \bibinfo{pages}{066804} (\bibinfo{year}{2006}).

\bibitem[{\citenamefont{M\"uller et~al.}(2004)\citenamefont{M\"uller, Heusler,
  Braun, Haake, and Altland}}]{kn:mueller2004}
\bibinfo{author}{\bibfnamefont{S.}~\bibnamefont{M\"uller}},
  \bibinfo{author}{\bibfnamefont{S.}~\bibnamefont{Heusler}},
  \bibinfo{author}{\bibfnamefont{P.}~\bibnamefont{Braun}},
  \bibinfo{author}{\bibfnamefont{F.}~\bibnamefont{Haake}}, \bibnamefont{and}
  \bibinfo{author}{\bibfnamefont{A.}~\bibnamefont{Altland}},
  \bibinfo{journal}{Phys. Rev. Lett.} \textbf{\bibinfo{volume}{93}},
  \bibinfo{eid}{014103} (\bibinfo{year}{2004}).

\bibitem[{\citenamefont{M\"uller et~al.}(2005)\citenamefont{M\"uller, Heusler,
  Braun, Haake, and Altland}}]{kn:mueller2005}
\bibinfo{author}{\bibfnamefont{S.}~\bibnamefont{M\"uller}},
  \bibinfo{author}{\bibfnamefont{S.}~\bibnamefont{Heusler}},
  \bibinfo{author}{\bibfnamefont{P.}~\bibnamefont{Braun}},
  \bibinfo{author}{\bibfnamefont{F.}~\bibnamefont{Haake}}, \bibnamefont{and}
  \bibinfo{author}{\bibfnamefont{A.}~\bibnamefont{Altland}},
  \bibinfo{journal}{Phys.\ Rev.\ E} \textbf{\bibinfo{volume}{72}},
  \bibinfo{eid}{046207} (\bibinfo{year}{2005}).

\bibitem[{\citenamefont{Richter and Sieber}(2002)}]{kn:richter2002}
\bibinfo{author}{\bibfnamefont{K.}~\bibnamefont{Richter}} \bibnamefont{and}
  \bibinfo{author}{\bibfnamefont{M.}~\bibnamefont{Sieber}},
  \bibinfo{journal}{Phys.\ Rev.\ Lett.} \textbf{\bibinfo{volume}{89}},
  \bibinfo{eid}{206801} (\bibinfo{year}{2002}).

\bibitem[{\citenamefont{Spehner}(2003)}]{kn:spehner2003}
\bibinfo{author}{\bibfnamefont{D.}~\bibnamefont{Spehner}}, \bibinfo{journal}{J.
  Phys. A} \textbf{\bibinfo{volume}{36}}, \bibinfo{pages}{7269}
  (\bibinfo{year}{2003}).

\bibitem[{\citenamefont{Turek and Richter}(2003)}]{kn:turek2003}
\bibinfo{author}{\bibfnamefont{M.}~\bibnamefont{Turek}} \bibnamefont{and}
  \bibinfo{author}{\bibfnamefont{K.}~\bibnamefont{Richter}},
  \bibinfo{journal}{J. Phys. A} \textbf{\bibinfo{volume}{36}},
  \bibinfo{pages}{L455} (\bibinfo{year}{2003}).

\bibitem{foot} Strictly speaking, one 
should also include trajectories for which transverse momenta
components are opposite, cf.\ Eq\ (\ref{eq:pW}). However in that case
the summation over trajectories contains a rapidly oscillating phase,
so that their net contribution to the sum vanishes in the classical
limit, see, {\em e.g.}, 
Refs.\ \onlinecite{kn:whitney2005,kn:jacquod2006,kn:rahav2006b}.

\bibitem[{\citenamefont{Rahav and Brouwer}(2006{\natexlab{a}})}]{kn:rahav2006b}
\bibinfo{author}{\bibfnamefont{S.}~\bibnamefont{Rahav}} \bibnamefont{and}
  \bibinfo{author}{\bibfnamefont{P.~W.} \bibnamefont{Brouwer}},
  \bibinfo{journal}{Phys. Rev. Lett.} \textbf{\bibinfo{volume}{96}},
  \bibinfo{pages}{196804} (\bibinfo{year}{2006}{\natexlab{a}}).

\bibitem[{\citenamefont{Tian et~al.}(2006)\citenamefont{Tian, Altland, and
  Brouwer}}]{kn:tian2006}
\bibinfo{author}{\bibfnamefont{C.}~\bibnamefont{Tian}},
  \bibinfo{author}{\bibfnamefont{A.}~\bibnamefont{Altland}}, \bibnamefont{and}
  \bibinfo{author}{\bibfnamefont{P.~W.} \bibnamefont{Brouwer}},
  \bibinfo{journal}{cond-mat/0605051}  (\bibinfo{year}{2006}).

\bibitem[{\citenamefont{Rahav and Brouwer}(2006{\natexlab{b}})}]{kn:rahav2006c}
\bibinfo{author}{\bibfnamefont{S.}~\bibnamefont{Rahav}} \bibnamefont{and}
  \bibinfo{author}{\bibfnamefont{P.~W.} \bibnamefont{Brouwer}},
  \bibinfo{journal}{unpublished}  (\bibinfo{year}{2006}{\natexlab{b}}).

\end{thebibliography}

\end{document}